\documentclass[journal]{IEEEtran}
\usepackage{amsfonts,amssymb,amsbsy,textcomp,marvosym,amsmath,caption,threeparttable,amsthm,subfigure}
\usepackage{eurosym,mathrsfs,fancyhdr,CJK,multicol,graphics,indentfirst,color,bm,upgreek,booktabs,graphicx}
\usepackage{bm}
\usepackage{graphicx}
\usepackage{tabularx}
\usepackage{multirow}
\usepackage{makecell}
\usepackage{subfigure}
\usepackage{algorithm}
\usepackage{algorithmic}
\usepackage{color}
\usepackage{enumitem}

\ifCLASSINFOpdf
\else
\fi
\begin{document}
%
\title{Joint Multi-grained Popularity-aware Graph Convolution Collaborative Filtering for Recommendation}


%

\author{Kang Liu,
	Feng Xue,
	Xiangnan He,
	Dan Guo,
	and Richang Hong,
	\thanks{(Corresponding author: Feng Xue.)} 
	\thanks{Kang Liu is with the School of Computer Science and Information Engineering, Key Laboratory of Knowledge Engineering with Big Data, Intelligent Interconnected Systems Laboratory of Anhui Province, Hefei University of Technology, Hefei, Anhui, China (e-mail: kangliu1225@gmail.com).}
	\thanks{Feng Xue is with the School of Software, Key Laboratory of Knowledge Engineering with Big Data, Intelligent Interconnected Systems Laboratory of Anhui Province, Hefei University of Technology, Hefei, Anhui, China (e-mail: feng.xue@hfut.edu.cn).}
	\thanks{Xiangnan He is with the University of Science and Technology of China, Hefei, Anhui, China (e-mail:  xiangnanhe@gmail.com).}
	\thanks{Dan Guo is with the School of Computer Science and Information Engineering, Key Laboratory of Knowledge Engineering with Big Data, Intelligent Interconnected Systems Laboratory of Anhui Province, Hefei University of Technology, Hefei, Anhui, China (e-mail: guodan@hfut.edu.cn).}
	\thanks{Richang Hong is with the School of Computer Science and Information Engineering, Key Laboratory of Knowledge Engineering with Big Data, Intelligent Interconnected Systems Laboratory of Anhui Province, Hefei University of Technology, Hefei, Anhui, China (e-mail: hongrc@hfut.edu.cn).}}


\markboth{Journal of \LaTeX\ Class Files,~Vol.~14, No.~8, August~2015}%
{Shell \MakeLowercase{\textit{et al.}}: Bare Demo of IEEEtran.cls for IEEE Transactions on Magnetics Journals}
%



\IEEEtitleabstractindextext{%
\begin{abstract}
Graph Convolution Networks (GCNs), with their efficient ability to capture high-order connectivity in graphs, have been widely applied in recommender systems. Stacking multiple neighbor aggregation is the major operation in GCNs. It implicitly captures popularity features because the number of neighbor nodes reflects the popularity of a node. However, existing GCN-based methods ignore a universal problem: users' sensitivity to item popularity is differentiated, but the neighbor aggregations in GCNs actually fix this sensitivity through Graph Laplacian Normalization, leading to suboptimal personalization.

In this work, we propose to model multi-grained popularity features and jointly learn them together with high-order connectivity, to match the differentiation of user preferences exhibited in popularity features. Specifically, we develop a \textbf{J}oint \textbf{M}ulti-grained \textbf{P}opularity-aware \textbf{G}raph Convolution \textbf{C}ollaborative \textbf{F}iltering model, short for JMP-GCF, which uses a popularity-aware embedding generation to construct multi-grained popularity features, and uses the idea of joint learning to capture the signals within and between different granularities of popularity features that are relevant for modeling user preferences. Additionally, we propose a multistage stacked training strategy to speed up model convergence. We conduct extensive experiments on three public datasets to show the state-of-the-art performance of JMP-GCF. The complete codes of JMP-GCF are released at https://github.com/kangliu1225/JMP-GCF.

\end{abstract}

}

\maketitle

\IEEEdisplaynontitleabstractindextext

%
\IEEEpeerreviewmaketitle

\section{Introduction}
%
%
%
%
\IEEEPARstart{P}{ersonalized} recommendation methods have been deployed in many applications \cite{music}\cite{social} to solve information overload and refine user experience in online services. Collaborative Filtering (CF) \cite{cf} is the mainstream algorithm for recommender systems because of its effectiveness and low computational overload. At its core is using historical user-item interactions to incorporate collaborative signals into the embedding process. 

However, CF suffers from the problem of sparsity. One solution is to enhance embeddings with auxiliary information, such as reviews \cite{review}, images \cite{visual1} \cite{vbpr}, social networks \cite{social}, knowledge graph\cite{kgcn} \cite{ripplenet}, and demographic characteristics \cite{svdfeature}; however, storing these additional data takes up too much memory, and most users resent the collection of personal information. Another solution is to consider neighbor information as an additional feature. For example, SVD++ \cite{svd++} and FISM \cite{fism} aggregate first-order neighbors  into user embeddings to enhance the modeling of user preferences. However, they suffer from two shortcomings: (1) the representation of item embedding is not enhanced, and (2) high-order interactions are not captured.

Graph Convolution Network (GCN) \cite{gcn}\cite{gat}\cite{graphsage} is an advanced deep learning technique for handling graph data that addresses the above two shortcomings. A common paradigm of GCN is to first process neighbor aggregation using a nonlinear neural network, and then iteratively perform this process to capture high-order neighbor information. 
GC-MC \cite{gcmc} is an early effort that uses a first-order GCN to aggregate neighbor information for all nodes in the user-item graph, thus enhancing both user embedding and item embedding. PinSAGE \cite{pinsage} and NGCF \cite{ngcf} extend the depth of GCN to capture important high-order interactions and hence achieve better performance. Inspired by SGCN \cite{sgc}, LR-GCCF \cite{lrgccf} and LightGCN\cite{light} further simplify the structure of GCN by removing the redundant nonlinear network layers in graph convolution, which both speeds up the model training and significantly improves performance. In summary, a linear GCN (which excludes nonlinear network layers) significantly outperforms a traditional nonlinear GCN for embedding generation in CF scenarios. 

In a user-item graph, the number of first-order neighbors directly reflects the popularity of the current node, and a linear GCN is essentially a process of iteratively performing neighbor aggregation. Therefore, a linear GCN implicitly incorporates popularity features into the node embedding. Additionally, the normalization performed after neighbor aggregation is equivalent to scaling the popularity features with a fixed granularity. Clearly, linear GCN-based methods that enhance the node embedding by high-order neighbor aggregation benefit from this implicit capture of popularity features.  
However, existing popularity-based methods\cite{popu1}\cite{popu2}\cite{popu3} perform poorly in recommendation tasks, mainly because they do not fully use interaction data to jointly learn CF signals and popularity features, and to achieve the complementation between them. Intuitively, a better solution is to use GCN as the base module to capture high-order CF signals, and incorporate popularity features into the embedding generation.

\begin{figure}
	\centering
	\includegraphics[height=6cm,width=8cm]{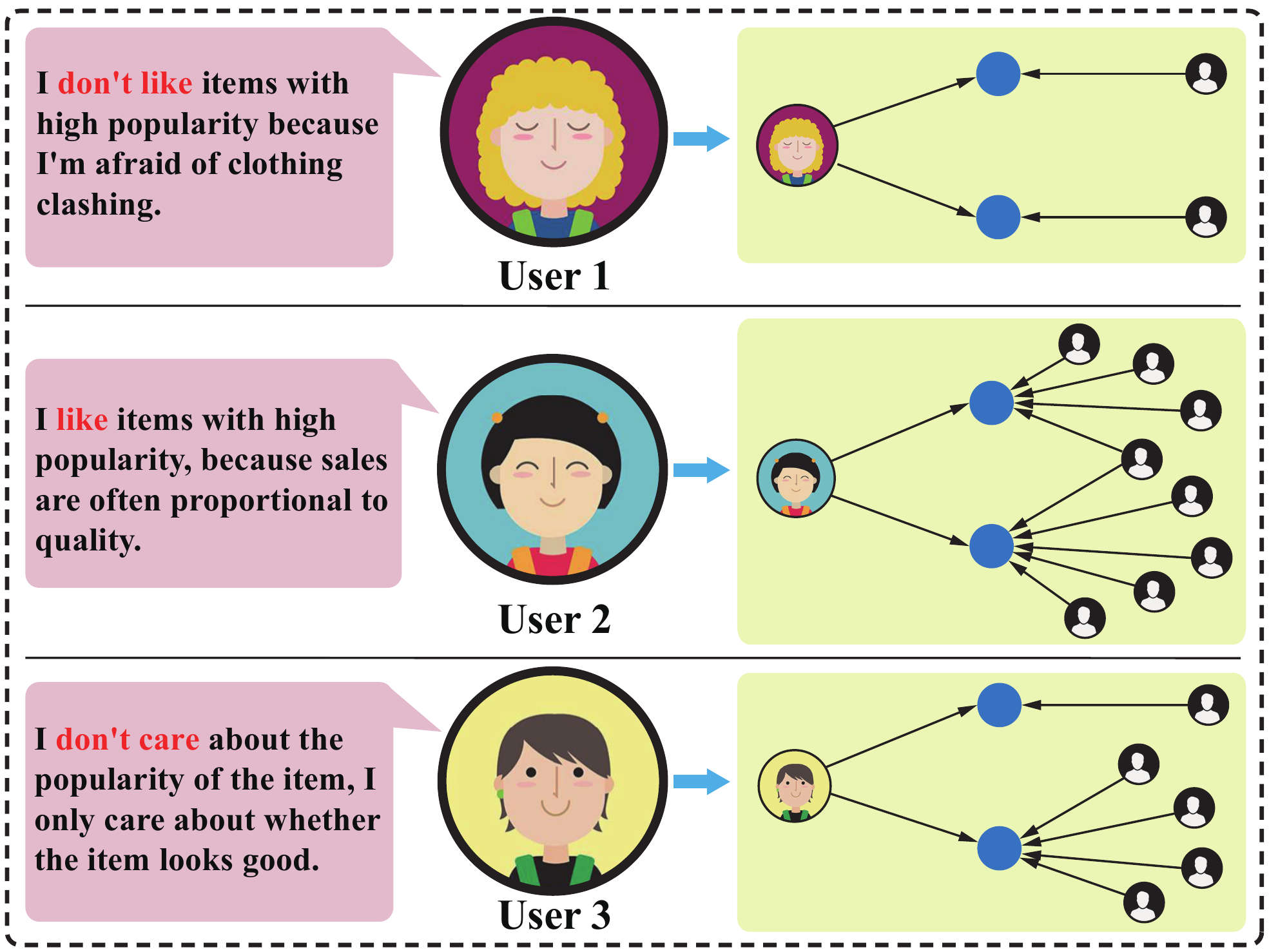}
	\caption{Illustration of the differentiation in user preferences exhibited in item popularity.  $User 1$ prefers items with low popularity; hence, the items that she buys have few interactions in the right-hand interaction graph;  $User 2$ likes items with high popularity; hence, she buys items that have many interactions; and $User 3$ is insensitive to the popularity of items, so she buys items with both high and low popularity.}
	\label{fig:motivation}
\end{figure}

Although neighbor aggregation can capture popularity features implicitly, \textbf{the sensitivity of user preferences to popularity is fixed, that is, neighbor aggregation captures popularity features at a fixed granularity.} This does not match the real situation, shown in Figure \ref{fig:motivation}, in which the sensitivity of users to popularity shows great differentiation. Existing methods ignore this inevitable problem, and therefore fail to achieve excellent personalization in respect of popularity. Thus, \textbf{capturing the distribution of users' preferences over popularity features with different granularities} is capable of modeling preferences better.

In this work, we propose a GCN-based model to achieve the above goal by jointly learning multi-grained popularity features and high-order interactions. Specifically, we propose a simplified GCN block, to capture high-order collaborative signals and construct multi-grained popularity-aware embeddings, and a layer selection mechanism to filter out the most expressive embeddings.
To achieve the joint learning of popularity features and collaborative signals, we propose separated Bayesian Personalized Ranking (BPR) loss to optimize the model and update parameters from the perspective of different layer semantics and different popularity granularities.
Additionally, we propose a multistage stacked training method to speed up convergence while facilitating the learning of popularity features.
Finally, we conduct extensive experiments on three million-size datasets to verify the effectiveness of our proposed JMP-GCF. 

The main contributions of our work are as follows:
\begin{itemize}
	\item  We highlight the critical importance of capturing multi-grained popularity features to match the differentiation of user preferences exhibited in item popularity. 
	\item We propose JMP-GCF, a novel GCN-based recommendation framework, which jointly learns multi-grained popularity features and high-order interactions to capture the signals related to modeling user preferences within and between different popularity granularities.
	\item We conduct extensive experiments on three public datasets. Experimental results demonstrate the state-of-the-art performance of  JMP-GCF. 
	\item To support the subsequent researches on JMP-GCF, we release the complete codes of JMP-GCF at https://github.com/kangliu1225/JMP-GCF.  
\end{itemize}

\section{METHODOLOGY}\label{sec:method}                                             
\begin{figure*}
	\centering
	\includegraphics[height=6.83cm,width=18cm]{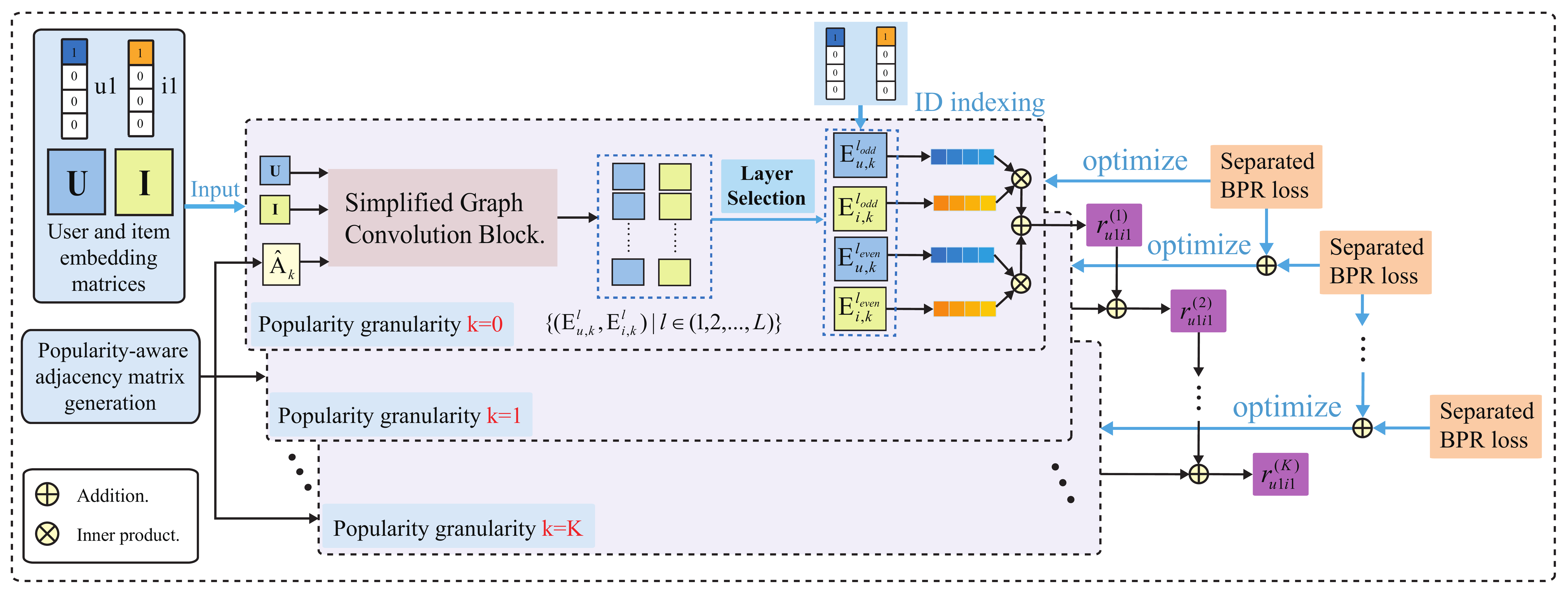}
	\caption{Illustration of JMP-GCF, where $k$ denotes the specific granularity in different simplified graph convolution blocks, and $K$ is the maximum popularity granularity. For a given user u1 and item i1, JMP-GCF is split into $K+1$ simplified graph convolution blocks to generate embeddings with different popularity granularities and optimal layers, and then the corresponding preference scores between u1 and i1 are calculated respectively, and finally, these embeddings are jointly optimized by using separated BPR loss so that multi-grained popularity features, layer semantics, and high-order connectivities are learned.  }
	\label{fig:model}
\end{figure*}

In this section, we present our proposed JMP-GCF model, whose general architecture is illustrated in Figure \ref{fig:model}. The model has four components: (1) the simplified graph convolution block, for integrating multi-grained popularity features into collaborative signals and generating embedding matrices; (2) the layer selection mechanism, for filtering out optimal graph convolution layers; (3) the model prediction\&optimizer layer to output predicted preference scores, and the separated BPR loss to optimize model parameters; and (4) the multistage stacked training strategy, for speeding up convergence and facilitating the learning of multi-grained popularity.
\subsection{Simplified Graph Convolution Block}\label{subsec:gcn}
\subsubsection{Embedding Generation}
\begin{figure}
	\centering
	\includegraphics[height=5.53cm,width=9cm]{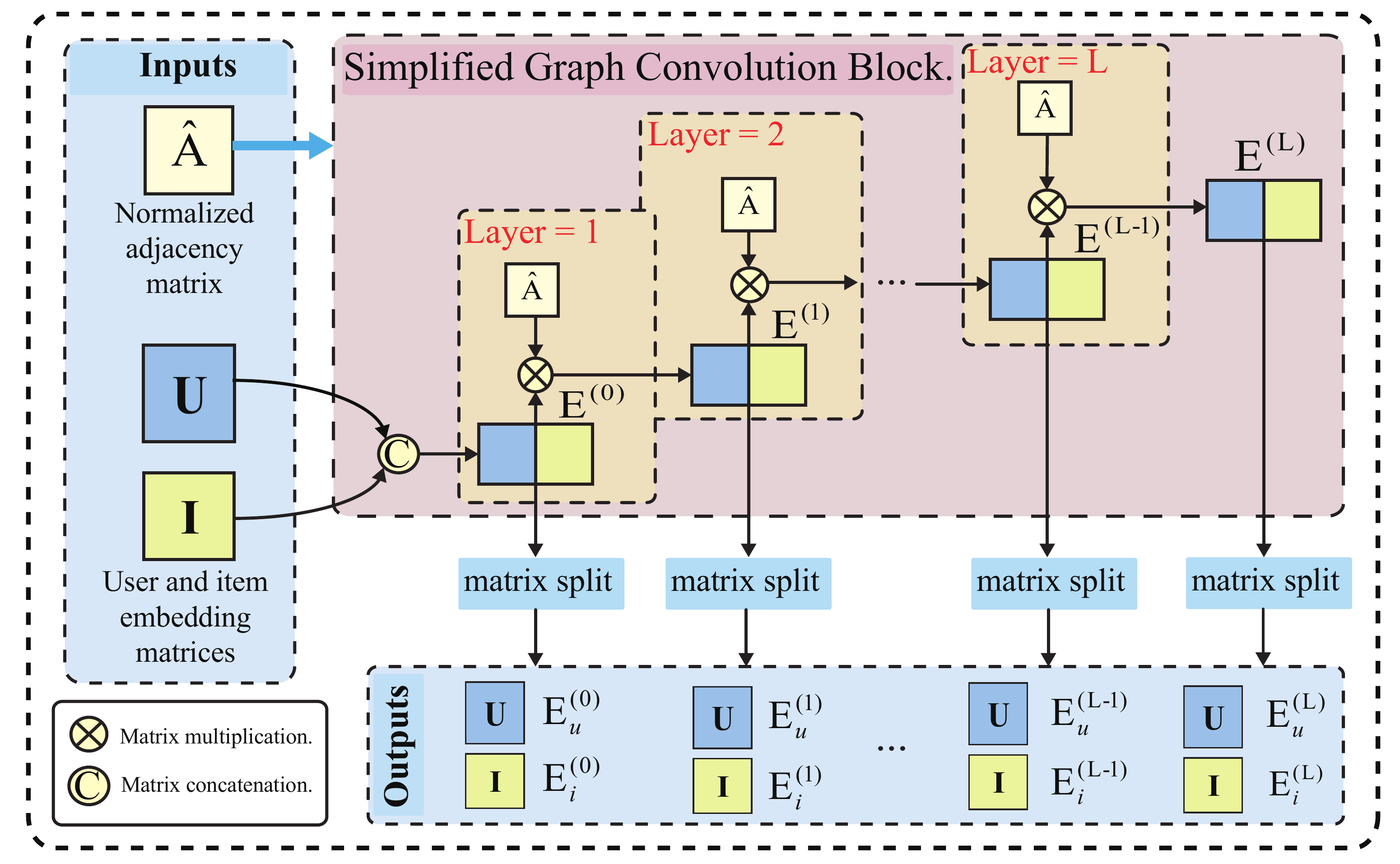}
	\caption{Illustration of the simplified graph convolution block. Given a normalized adjacency matrix $\hat{\textbf A}$, and randomly initialized user and item embedding matrices, $\textbf{U}$ and ${\textbf I} $, this block outputs the embedding matrices of users and items obtained at each graph convolution layer.}
	\label{fig:rgcn}
\end{figure}

According to prior work \cite{light}, the network layers and nonlinear activation functions in traditional GCNs are redundant in CF scenarios. Therefore, we construct an embedding generation block that uses a  simplified graph convolution to implement neighbor aggregation and message passing, named simplified graph convolution block. Figure \ref{fig:rgcn} shows the detailed structure of  this block.
The first graph convolution operation, that is, the first yellow rectangle (marked $Layer=1$) in Figure \ref{fig:rgcn}, can be represented by Equation \ref{eq:e1}:
\begin{equation}\label{eq:e1}
\textbf E^{(1)} = {(\bf D+\bf I)}^{- \frac{1}{2}} {(\bf A+\bf I)} {(\bf D+\bf I)}^{-\frac{1}{2}} \times \textbf [\textbf U || \textbf I ],
\end{equation}
where $\textbf A \in \mathbb{R}^{(m+n)\times(m+n)}$ is the adjacency matrix, $\textbf D $ is the degree matrix of $\textbf A$, $\bf{I}$ is the identity matrix of $\bf{A}$, $\textbf U \in \mathbb{R}^{m\times k}$ and $ \textbf I \in \mathbb{R}^{n \times k}$ are the randomly initialized user and item embedding matrices, respectively, $m$ and $n$ denote the number of users and items, respectively, $k$ is the dimension length of the embedding, $||$ represents matrix concatenation, and $\textbf E^{(1)} \in \mathbb{R}^{(m+n)\times k}$ is the output of the first graph convolution layer. 

Because the graph convolution operation is stackable in this block, assuming that the current layer is $l$, that is, the $l$-th yellow rectangle, the output of this layer can be formulated as follows:

\begin{equation}\label{eq:el}
\textbf E^{(l)}={(\bf D+\bf I)}^{-\frac{1}{2}} {(\bf A+\bf I)} {(\bf D+\bf I)}^{-\frac{1}{2}} \times \textbf E^{(l-1)},
\end{equation}
where $\textbf E^{(l)}$ and $\textbf E^{(l-1)}$ denote the embedding matrices obtained at layers $l$ and $l-1$, respectively. Note that the embedding matrices output at each layer can be redivided into user and item embedding matrices. Setting the total number of graph convolution layers to $L$, and the outputs of this block are as follows:

\begin{equation}
\label{eq:matrix}
\{(\textbf E_u^{(l)}, \textbf E_i^{(l)})|l \in (1,2,...,L)\},
\end{equation}
where $\textbf E_u^{(l)}$ and $\textbf E_i^{(l)}$ are the embedding matrices obtained at the $l$-th layer of this block. 

\subsubsection{Popularity-aware Embedding Generation}\label{sec:sgcb2}
Intuitively, popularity is an important component of user preferences, and users have different sensitivities to popularity features. Therefore, the weights of nodes with high popularity (or degree) should be appropriately increased to improve their sensitivity to popularity.
We can obtain a normalized adjacency matrix containing popularity features at different granularity levels by fine-tuning the exponent of the degree matrix as follows:
\begin{equation}\label{eq:a}
\hat{\textbf A_k} =  {(\bf D+\bf I)}^{-\frac{1}{2}}  {(\bf A+\bf I)}  {(\bf D+\bf I)}^{-\frac{1}{2} + kc},
\end{equation}
where $c$ is a constant, which is the smallest unit by which the granularity of popularity can vary (we set $c$ to 0.1 in our experiments), and we set $k$ to the granulairty of popularity, which indicates how much the embeddings varies according to popularity. A greater value of $kc$, represents a higher weight for the node popularity features and forces the model to recommend items with higher popularity to users. Specifically, the embedding generation process of Equation \ref{eq:a} can be viewed as the simultaneous capture of the higher-order CF signals and the popularity feature with granularity $k$. When $k$=0, Equation \ref{eq:a} is equivalent to the traditional graph convolution operation ($cf.$ \cite{gcn}), $i.e.$, only the modeling of the CF signal is considered. when $k$ = 1, the model starts to incorporate small granularity of popularity features and amplifies the value of the embeddings appropriately based on that granularity $k$ = 1. When $k$ = 2 or larger, the model incorporates a larger popularity granularity and scales up the embedding values more, which corresponds to the model being more sensitive to popularity. Since users have different sensitivities to popularity features, we set different popularity granularity $k$ in embedding generation so that the recommendation results of the model can cover items with different popularity as much as possible, thus improving user satisfaction. 

Inputting $\hat{\textbf A_k}$ to the simplified graph convolution block as a normalized adjacency matrix, we obtain the embedding matrices at the $l$-th graph convolution layer as follows:
\begin{equation}\label{eq:ake}
\textbf E_k^{(l)}=\hat{\textbf A_k} \times \textbf E_k^{(l-1)},
\end{equation}
where $\textbf E_k^{(0)}=\textbf [\textbf U || \textbf I ]$, and $\textbf E_k^{(l)}$ can be redivided into user and item embedding matrices. We obtain the following user and item embedding matrices at each layer:
\begin{equation}\label{eq:akeo}
\{(\textbf E_{u,k}^{(l)}, \textbf E_{i,k}^{(l)})|l \in (1,2,...,L), k \in (0,1,...,K)\},
\end{equation}
where $K$ denotes the maximum popularity granularity. In our work, for simplicity, we set $K$ to 2 to obtain three sets of user and item embedding matrices at each layer.

Having described how to incorporate the multi-grained popularity features into the embedding generation, we next introduce a straightforward strategy that has the potential to better personalize the matching of user preferences and different popularity levels. Specifically, we assign corresponding weights $\lambda_k$ to the user embedding matrices at different popularity granularities $k$. We formulate this process as follows:
\begin{equation}
\{\textbf E_{u,k}^{(l)*} = \lambda_k \cdot \textbf E_{u,k}^{(l)}|l \in (1,2,...,L), k \in (0,1,...,K)\},
\end{equation}
where $\lambda_k$ represents the user's preference level for different granularities of popularity, which can be obtained in two ways. One is to construct an attention network to calculate the sensitivity of user $u$ to different popularity granularities. we leave it for future study. The other is that in an online environment, users set the granularity value according to their own needs, thus controlling the popularity of the tweeted content.
Specifically, all $\lambda_k$ are set to 1 by default, and if the user chooses a specific granularity value $k$ = 1, we make $\lambda_1$ = 10 (or other value greater than 1), which is equivalent to that the system will be more inclined to generate recommendations based on that popularity granularity. In this work, we only briefly introduce this strategy without experimentally verifying it, since it requires real-time user participation.


\begin{figure}
	\centering
	\includegraphics[height=2.5cm,width=9cm]{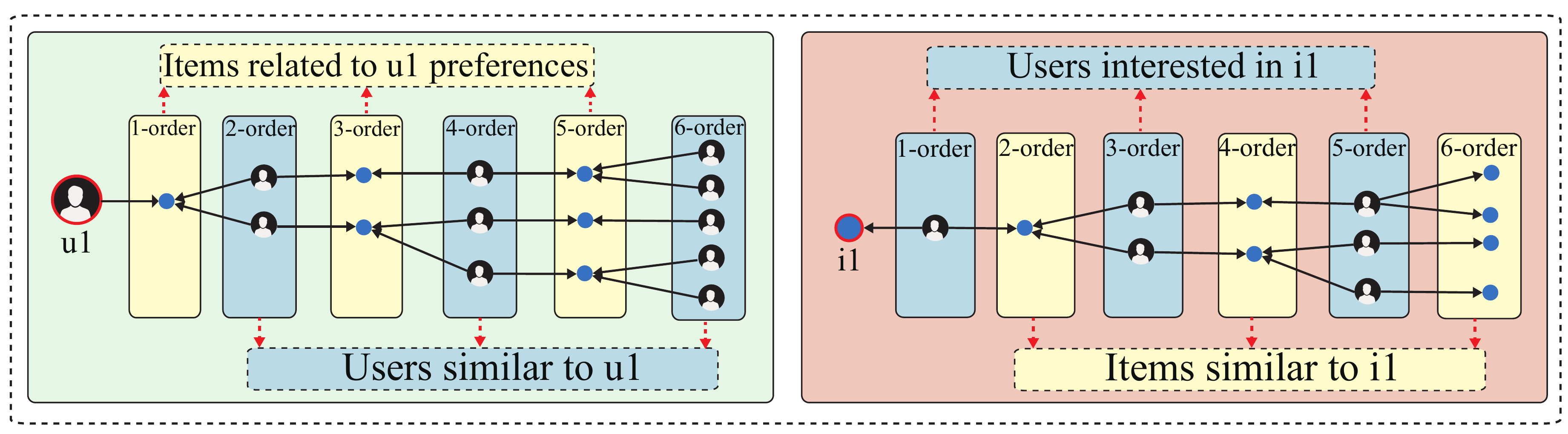}
	\caption{Illustration of the relationship between the semantics of neighbor nodes and the number of hops where the neighbor nodes are located, $u_1$ and $i_1$ are the target user and item, respectively.}
	\label{fig:semantic}
\end{figure}
\begin{algorithm}
	\raggedright           
	\caption{: LAYER SELECTION}
	\label{alg:ls}      
	\textbf{Input:} Dataset $\bf D$; threshold parameter $\alpha$; neighborhood function to compute the number of $k$-hop neighbors for current node $u$, $N^k(u)$.\\  
	\textbf{Output:} Optimal odd layer $l_{odd}$; optimal even layer $l_{even}$. \\ 
	\begin{algorithmic}[1]
		\STATE Compute the number of total users {and items in $\bf D$, $N_{user}$ and $N_{item}$, respectively}.
		\STATE Randomly sample 100 users into a set, $H_{user}$.
		\STATE Let $l_{odd} \leftarrow 0$, $l_{odd} \leftarrow 0$, $k_1 \leftarrow 0$, $k_2 \leftarrow 0$.
		\WHILE{$l_{odd} = 0$}
		\STATE {Obtain next odd layer,} $k_1 \leftarrow 2 \times  k_1 + 1$. 
		\STATE {Initialize the total number of odd layer nodes,} $S_{odd} \leftarrow 0$.
		\FOR {$u \in  H_{user}$}
		\STATE $S_{odd} \leftarrow S_{odd} + N^{k_1}(u)$
		\ENDFOR
		\IF {$(S_{odd} / N_{item})/100 >= \alpha$}
		\STATE {Obtain optimal odd layer,} $l_{odd} \leftarrow k_1$
		\ENDIF
		\ENDWHILE
		
		\WHILE{$l_{even} = 0$}
		\STATE {Obtain next even layer,} $k_2 \leftarrow 2 \times  k_2 + 2$.
		\STATE {Initialize the total number of even layer nodes,} $S_{even} \leftarrow 0$.
		\FOR {$u \in  H_{user}$}
		\STATE $S_{even} \leftarrow S_{even} + N^{k_2}(u)$
		\ENDFOR
		\IF {$(S_{even} / N_{user})/100 >= \alpha$}
		\STATE {Obtain optimal even layer,} $l_{even} \leftarrow k_2$
		\ENDIF
		\ENDWHILE
		
		\STATE \textbf{return} $l_{odd}$ and $l_{even}$
	\end{algorithmic}
	\label{ag:1}
\end{algorithm}
\subsection{Layer Selection Mechanism}\label{subsec:lsm}
In this section, we propose a layer selection mechanism to ensure the expressiveness of embedding. This mechanism filters out the optimal graph convolution layer from the simplified graph convolution block. 

In the simplified graph convolution block, we argue that the embedding output at layer $l$ contains the complete information of the previous $l-1$ layers, that is, the embedding at a higher layer has a larger interaction space (the ratio of incorporated neighbor nodes to global nodes), and is thus more expressive. The following derivation proves the above assumption when $l=3$ (for simplicity, normalization is not considered here):
\begin{equation}
\begin{split}
\textbf{E}^{(3)} = &(\textbf{A}+\textbf{I})\textbf{E}^{(2)} =
(\textbf{A}+\textbf{I})(\textbf{A}+\textbf{I})\textbf{E}^{(1)} \\
= &\textbf{A}(\textbf{A}+\textbf{I})\textbf{E}^{(1)} + (\textbf{A}+\textbf{I})\textbf{E}^{(1)}
=\textbf{A}(\textbf{A}+\textbf{I})\textbf{E}^{(1)} + \textbf{E}^{(2)},
\end{split}
\end{equation}
where $\textbf{E}^{(3)}$, $\textbf{E}^{(2)}$, and $\textbf{E}^{(1)}$ denote the embedding matrices output at the third, second, and first layers, respectively. 

It is worth mentioning that too high a layer tends to introduce substantial noise (that is, it contains many nodes that are completely unrelated to the current node). Therefore, it is important to choose the appropriate graph convolution layers to ensure sufficient interaction space and avoid the contamination by noise nodes. This contrasts with the layer aggregation mechanism\cite{ngcf}\cite{lrgccf}\cite{light}, which uses the embeddings of all the graph convolution layers.

In addition to considering the size of interaction space, we believe that the optimal graph convolution layers can sufficiently capture two layer semantics (heterogeneous and homogeneous nodes), which are interpreted in Figure \ref{fig:semantic}. 
Because these two layer semantics have different interaction space sizes in each layer, the positive effects of both semantics can be maximized if their interaction space is sufficiently large. For this reason, we obtain the optimal odd layer $l_{odd}$ and even layer $l_{even}$ output by Algorithm \ref{alg:ls}, which ensures that the interaction spaces of the two semantics are both sufficient.

For a specific dataset, we can determine whether the current hop corresponds to a semantically sufficient graph convolution layer by calculating the interaction space size of the two semantics under each hop. In Algorithm \ref{alg:ls}, $S_{odd} / N_{item} $ denotes the interaction space size for the first semantic at hop $k_1$. Similarly, $S_{even} / N_{user}$ denotes the interaction space size for the other semantic at hop $k_2$. We empirically set the threshold parameter $\alpha$ to 50\% for both semantics. Finally, the algorithm outputs $l_{odd}$ and $l_{even}$, which are the odd and even layers for which the interaction space size first reaches $\alpha$.

\subsection{Model Prediction \& Optimization.}\label{sec:preopt}
According to the simplified graph convolution block and layer selection mechanism, we obtain the user and item embedding matrices that contain sufficient layer semantics and multi-grained popularity features as follows:
\begin{equation} \label{eq:H}
\{(\textbf E_{u,k}^{(l)}, \textbf E_{i,k}^{(l)})|l \in (l_{odd}, l_{even}), k \in (1, 2,..., K)\},
\end{equation}
where $\textbf{E}_{u,k}$ and $\textbf{E}_{i,k}$ denote the user and item embedding matrices at the popularity granularity of $k$, respectively. 

The embedding representations of target users and items can be obtained directly by indexing the IDs ($u_1$ and $i_1$) in these matrices, as follows:
\begin{equation}
\{(\textbf e_{u_1,k}^{(l)}, \textbf e_{i_1,k}^{(l)})|l \in (l_{odd}, l_{even}), k \in (1,2,...,K)\},
\end{equation}
where $\textbf e_{u_1,k}$ and $\textbf e_{i_1,k}$ denote the embedding representations for user $u_1$ and item $i_1$ at the popularity granularities of $k$, respectively.

We use a simple inner product operation to calculate the prediction scores at multiple popularity granularities and two layer semantics, and sum them to obtain the final prediction score between the target user and item. The specific prediction function is as follows:
\begin{equation}\label{eq:pred}
\begin{split}
r_{u_1i_1}& = \sum_{k=0}^{K} [{(\textbf{e}_{u_1,k}^{l_{odd}})}^T\textbf{e}_{i_1,k}^{l_{odd}} + {(\textbf{e}_{u_1,k}^{l_{even}})}^T\textbf{e}_{i_1,k}^{l_{even}}].
\end{split}
\end{equation}

\noindent\textbf{Separated BPR Loss.} To facilitate the learning of layer semantics and popularity features for user preferences, we propose a separated BPR loss to optimize the model parameters as follows:
\begin{equation}\label{eq:loss}
\begin{split}
L = &\sum_{(u,i,j)\in O} \sum_{k=0}^{K}\{-ln\sigma[(\textbf{e}_{u,k}^{l_{odd}})^T\textbf{e}_{i,k}^{l_{odd}}  -  (\textbf{e}_{u,k}^{l_{odd}})^T\textbf{e}_{j,k}^{l_{odd}}]- \\&
ln\sigma[(\textbf{e}_{u,k}^{l_{even}})^T\textbf{e}_{i,k}^{l_{even}}  -  (\textbf{e}_{u,k}^{l_{even}})^T\textbf{e}_{j,k}^{l_{even}} ] \}+ \lambda ||\textbf{H}||_2^2,
\end{split}
\end{equation}
where $O=\{(u,i,j) \mid i\in N_u, j \notin N_u \}$ denotes the training data, $N_u$ denotes the observed item set for user $u$, and $\sigma(\cdot)$ is the sigmoid function. We apply $L_2$ regularization on $\textbf H$ controlled by $\lambda$, where  $\textbf H$ denotes the embedding matrices in Equation \ref{eq:H}.

Without considering the popularity granularity ($k$), we let ${d}_{ui\leftarrow uj} = (\textbf{e}_{u})^T\textbf{e}_{i}  -  (\textbf{e}_{u})^T\textbf{e}_{j}$, and further derive the core term in Equation \ref{eq:loss} as follows:
\begin{equation}
\begin{split}
&-ln\sigma({d}_{ui\leftarrow uj}^{l_{odd}}) -  ln\sigma({d}_{ui\leftarrow uj}^{l_{even}} ) =  \\&-ln[\frac{1}{1+e^{-{d}_{ui\leftarrow uj}^{l_{odd}}}+e^{-{d}_{ui\leftarrow uj}^{l_{even}}} + e^{-({d}_{ui\leftarrow uj}^{l_{odd}} + {d}_{ui\leftarrow uj}^{l_{even}})}}].
\end{split}
\end{equation}

The above expanded formulas illustrate that the target of our proposed loss function is to essentially maximize the ${d}_{ui\leftarrow uj}^{l_{odd}}$, ${d}_{ui\leftarrow uj}^{l_{even}}$, and $({d}_{ui\leftarrow uj}^{l_{odd}} + {d}_{ui\leftarrow uj}^{l_{even}})$, that is, to maximize the distance between positive and negative samples when considering the two layer semantics both individually and jointly.

This derivation demonstrates that the separated BPR loss can explicitly capture signals related to modeling user preferences within and between different semantics.
In addition, this joint learning of separated BPR loss also considers the effects of different granularities of popularity features simultaneously, so that features within and interactions between different granularities are also successfully captured and incorporated into the process of modeling user preferences.
We verify the effectiveness of separated BPR loss in Section \ref{ex:subsubsec:loss}.

\subsection{Multistage Stacked Training}\label{subsec:mst}
To effectively incorporate multi-grained popularity features into the embedding generation and speed up the training convergence, we propose a multistage stacked training method, which divides the overall framework into multiple training phases depending on the granularity of popularity features, as shown in Figure \ref{fig:model}. In the following, we provide details of the model optimization for each training phases. It is worth noting that the model in each training phase can perform the recommendation task independently.
\subsubsection{\textbf{First Training Phase}}
We first consider the learning of coarse-grained popularity features because it can accelerate the convergence of model training (evidences in Section \ref{sec:rq3}). 
According to Eqs. \ref{eq:a}, \ref{eq:ake}, and \ref{eq:akeo} and the layer selection mechanism, the embedding matrices with maximum popularity granularity are $\{(\textbf E_{u,K}^{(l)}, \textbf E_{i,K}^{(l)})|l \in (l_{odd}, l_{even})\}$, where $K$ is the maximum increase in the  magnitude of popularity granularity. Using the ID indexing, we obtain the embedding corresponding to the user and item as $\{(\textbf e_{u,K}^{(l)}, \textbf e_{i,K}^{(l)})|l \in (l_{odd}, l_{even})\}$.
We use the following formulas to compute the preference score between the target user $u_1$ and item $i_1$ :                                      
\begin{equation}\label{eq:pred1}
{r_{u1i1}^{(1)}}={(\textbf{e}_{u1,K}^{l_{odd}})}^T \textbf{e}_{i1,K}^{l_{odd}} + {(\textbf{e}_{u1,K}^{l_{even}})}^T \textbf{e}_{i1,K}^{l_{even}}.
\end{equation}
Additionally, we use the following separated BPR loss to optimize the model parameters:
\begin{equation}
\begin{split}\label{eq:loss1}
L_1 = &\sum_{(u,i,j)\in O}-ln\sigma\left[{(\textbf{e}_{u,K}^{l_{odd}})}^T \textbf{e}_{i,K}^{l_{odd}}-{(\textbf{e}_{u,K}^{l_{odd}})}^T \textbf{e}_{j,K}^{l_{odd}}\right]- \\&ln\sigma \left[{(\textbf{e}_{u,K}^{l_{even}})}^T \textbf{e}_{i,K}^{l_{even}}-{(\textbf{e}_{u,K}^{l_{even}})}^T \textbf{e}_{j,K}^{l_{even}} \right]+\lambda ||\textbf{H}_1||_2^2,
\end{split}
\end{equation}
where $\textbf{H}_1=\{(\textbf E_{u,K}^{(l)}, \textbf E_{i,K}^{(l)})|l \in (l_{odd}, l_{even})\}$ denotes the trainable embedding representations in layers ${l_{odd}}$ and ${l_{even}}$ in the current training phase.

\subsubsection{\textbf{Stacked Training Phase}}
When the first training phase has sufficiently captured coarse-grained popularity features, we focus on learning finer-grained features in subsequent training, and then stack these features on top of the previous training to achieve joint learning of multi-grained features and collaborative signals. 
Assuming that the current training phase is $p$, we obtain the embedding matrices at the corresponding popularity granularity are $\{(\textbf E_{u,k}^{(l)}, \textbf E_{i,k}^{(l)})|l \in (l_{odd}, l_{even})\}$, where $k = K-p+1$.
The embedding representations for users and items from these embedding matrices are $\{(\textbf e_{u,k}^{(l)}, \textbf e_{i,k}^{(l)})|l \in (l_{odd}, l_{even})\}$.
Additionally, the prediction and objective functions of this phase can be seen as a stack on top of the previous training phase. 

The prediction function in the current training phase $p$ is as follows:
\begin{equation}\label{eq:pred2}
\begin{split}
{r_{u1i1}^{(p)}}={(\textbf{e}_{u1,k}^{l_{odd}})}^T \textbf{e}_{i1,k}^{l_{odd}} + {(\textbf{e}_{u1,k}^{l_{even}})}^T \textbf{e}_{i1,k}^{l_{even}} + {r_{u1i1}^{(p-1)}},
\end{split}
\end{equation}
where ${r_{u1i1}^{(p)}}$ is the predicted score for user $u_1$ and item $i_1$ in the current training phase, and ${r_{u1i1}^{(p-1)}}$ is the predicted score in the $(p-1)$-th training phase.
We use the following separated BPR loss to optimize the model parameters for the current training phase:
\begin{equation}
\begin{split}\label{eq:lossk}
&L_{p}= \sum_{(u,i,j)\in O}-ln\sigma[{(\textbf{e}_{u,k}^{l_{odd}})}^T \textbf{e}_{i,k}^{l_{odd}}-{(\textbf{e}_{u,k}^{l_{odd}})}^T \textbf{e}_{j,k}^{l_{odd}}] - \\&ln\sigma [{(\textbf{e}_{u,k}^{l_{even}})}^T \textbf{e}_{i,k}^{l_{even}}-{(\textbf{e}_{u,k}^{l_{even}})}^T \textbf{e}_{j,k}^{l_{even}} ]+\lambda ||\textbf{H}_p||_2^2 + L_{p-1},
\end{split}
\end{equation}
where $L_{p}$ is the optimization target of the current training phase of the model, $L_{p-1}$ is the optimization target of the $(p-1)$-th training phase, and $\textbf{H}_p = \{(\textbf E_{u,k}^{(l)}, \textbf E_{i,k}^{(l)})|l \in (l_{odd}, l_{even})\}$ denotes all trainable embedding representations in the current phase.
It is worth noting that Eqs. \ref{eq:lossk} and \ref{eq:loss} are equivalent when the training phase $p$ is $K+1$.

\section{EXPERIMENT}\label{sec:experiment}
We conduct experiments to evaluate our proposed JMP-GCF, and further experimental analysis to verify the effectiveness of each component of JMP-GCF. We aim to answer the following questions:
\begin{itemize} 
	\item \textbf{RQ1}: How does JMP-GCF perform compared with other state-of-the-art recommender methods?
	\item \textbf{RQ2}: Are all components (layer selection, separated BPR loss, popularity integration, multistage stacked training) of JMP-GCF helpful?
	\item \textbf{RQ3}: How does the joint learning of multi-grained popularity features and high-order interactions affect the JMP-GCF?
\end{itemize}

\subsection{Experimental Settings}
\subsubsection{\textbf{Datasets}} For a fair comparison, we conduct experiments on the three public datasets, Gowalla \cite{gowalla}, Yelp2018, and Amazon-Book \cite{visual1}, used in \cite{ngcf} \cite{light}. We present the details of these three datasets in Table \ref{tab:statistics}. We randomly sample 80\% of the interaction data for each user as the training set and use the remaining 20\% as the test set. Furthermore, we sample 10\% of the interaction data from the training set as the validation set to tune the hyper-parameters. It is worth mentioning that the datasets and dataset settings we use are the same as those of the previous GCN-based works, $i.e.$, LightGCN \cite{light} and NGCF\cite{ngcf}.

\renewcommand{\arraystretch}{1.1}
\renewcommand\tabcolsep{3.0 pt} 
\begin{table}
	\centering
	\caption{Statistics of the datasets.}
	\label{tab:statistics}
	\begin{tabular}{c|c|c|c|c}
		\hline
		\bf{Dataset}&\bf{User \#}&\bf{Item \#}&\bf{Interaction \#}&\bf{Density}\cr
		\hline
		\hline
		\bf{Gowalla}&29,858&40,981&1,027,370&0.00084\cr
		\hline
		\bf{Yelp2018}&31,668&38,048&1,561,406&0.00130\cr
		\hline
		\bf{Amazon-Book}&52,643&91,599&2,984,108&0.00062\cr
		\hline
	\end{tabular}
\end{table}

\subsubsection{\textbf{Evaluate Metrics}} We select two protocols \cite{ngcf}: $Recall@20$ and $NDCG@20$ which are commonly used in recent works \cite{ngcf} \cite{kgat} \cite{light} to evaluate model performance. Specifically, we compute the average $Recall@20$ and $NDCG@20$ for each user in the test set. Note that for each user, we used all the items that the user had not interacted with as negative samples.

\subsubsection{\textbf{Baselines}} To demonstrate the effectiveness of our proposed JMP-GCF, we compare it with the following baselines:
\begin{itemize} 
	\item \textbf{BPRMF}\cite{mf}: This is a classical matrix factorization (MF) method with BPR loss, which is widely used as a recommendation baseline.
	\item \textbf{SVD++}\cite{svd++}: This is a variant of MF, which integrates the historical interactions into user embeddings. It can also be regarded as a one-layer GCN that only passes messages for user nodes.
	\item \textbf{NeuMF} \cite{ncf}: This is a neural CF method, which uses a nonlinear neural networks as an interaction function instead of a simple inner product of MF. 
	\item \textbf{GC-MC} \cite{gcmc}: This is a GCN-based encoder-decoder framework, which only uses one nonlinear graph convolution layer to capture first-order interactions for users and items. 
	\item \textbf{NGCF} \cite{ngcf}: This method adopts a GCN technique with multiple nonlinear graph convolution layers, and it concatenates the embeddings obtained at each layer as the final representations of users and items. 
	\item \textbf{LR-GCCF}\cite{lrgccf}: This model fine-tunes the structure of NGCF by removing the nonlinear activation function in the graph convolution layer.
	\item \textbf{LightGCN} \cite{light}:  This is a GCN-based CF model, which generates the embeddings with a simplified GCN and uses the summation of the embeddings obtained at each layer as the final representation. 
	\item {\textbf{SGL} \cite{sgl}: This is the latest GCN-based CF model, which employs node dropout, edge dropout, and random walk to construct different graph views, and uses self-supervised contrastive learning to maximize the agreement between different views. Additionally, SGL applies LightGCN as the base module for embedding generation.}
\end{itemize}

\subsubsection{\textbf{Hyper-parameter Settings}}
For all methods of comparison, we set the embedding size and batch size to 64 and 2048, respectively. We tune the learning rate in \{$10^{-4}$, $10^{-3}$, $10^{-2}$, $10^{-1}$\} and search the coefficient of $L_2$ regularization in \{$10^{-6}$, $10^{-5}$, $10^{-4}$, $10^{-3}$, $10^{-2}$, $10^{-1}$\}. Specifically, $10^{-3}$ is the optimal learning rate for the three datasets. $10^{-4}$, $10^{-4}$, and $10^{-5}$ are the optimal coefficients of $L_2$ regularization for the Gowalla, Yelp2018, and Amazon-Book, respectively. Besides, we use the Xavier initializer \cite{xavier} to achieve the embedding initialization for all models.

\renewcommand{\arraystretch}{1.3}
\renewcommand\tabcolsep{4 pt} 
\begin{table}
	\centering
	\caption{Overall performance comparison.}
	\label{tab:overall}
	\begin{tabular}{|c|c|c|c|c|c|c|}
		\hline
		\multirow{2}{*}{{{Method}}}&
		\multicolumn{2}{c|}{{{Gowalla}}}&\multicolumn{2}{c|}{{{Yelp2018}}}&\multicolumn{2}{c|}{{{Amazon-Book}}}\cr\cline{2-7}
		&{recall}&{ndcg}&{recall}&{ndcg}&{recall}&{ndcg}\cr
		\hline
		\hline
		{BPRMF}&0.1291&0.1109&0.0433&0.0354&0.0250&0.0196\cr
		{SVD++}&0.1439&0.1297&0.0500&0.0412&0.0332&0.0251\cr
		{NeuMF}&0.1399&0.1212&0.0451&0.0363&0.0258&0.0200\cr
		\hline
		{GC-MC}&0.1395&0.1204&0.0462&0.0379&0.0288&0.0224\cr
		{NGCF}&0.1569&0.1327&0.0579&0.0477&0.0337&0.0261\cr
		\hline
		{LR-GCCF}&0.1701&0.1452&0.0604&0.0498&0.0375&0.0296\cr
		{LightGCN}&0.1830&0.1554&0.0649&0.0530&0.0411&0.0315\cr
		 {SGL}& {0.1810}& {0.1531}& {0.0675}& {0.0555}& {0.0478}& {0.0379}\cr
		\hline
		\textbf{JMP-GCF}&{\textbf{0.1871}}&{\textbf{0.1583} }&{\textbf{ 0.0702}}&{\textbf{0.0577} }&{\textbf{0.0499} }&{ \textbf{0.0388}}\cr
		\hline
		\hline
		\%Improv.&2.24\%&1.87\%&4.00\%&3.96\%&4.39\%&2.37\%\cr\hline
	\end{tabular}
\end{table}
\begin{figure}
	\centering
	\includegraphics[height=8.5cm,width=8.5cm]{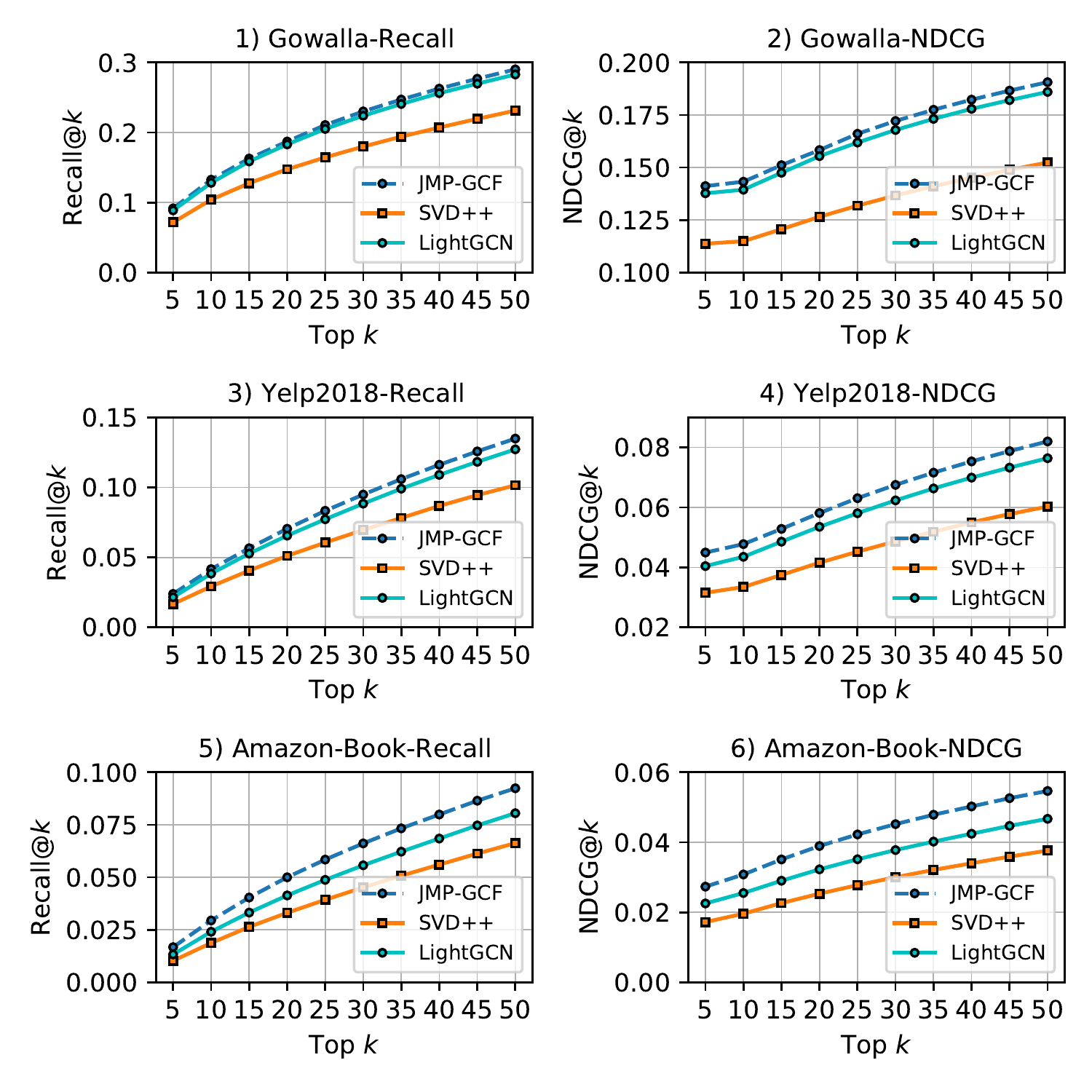}
	\caption{Performance comparison of JMP-GCF, SVD++, and LightGCN $w.r.t.$ different recommendation list size $k$ on the Gowalla, Yelp2018, and Amazon-Book datasets.}
	\label{fig:topk}
\end{figure}
\subsection{Overall Comparison (RQ1)}
Table \ref{tab:overall} reports the performance of all the methods. We obtain the following findings:
\begin{itemize} 
	\item BPRMF performs worst in all cases, which indicates that a simple inner product fails to capture complex interactions between nodes in the user-item graph. NeuMF outperforms BPRMF across all datasets, which verifies that using nonlinear neural networks can distill complex and nonlinear connectivities between nodes.
	
	\item GC-MC outperforms NeuMF on the Yelp2018 and Amazon-Book datasets, which verifies that using first-order GCN to incorporate one-hop neighbors can improve model performance. However, GC-MC is slightly worse than NeuMF on the Gowalla dataset, which might be because the nonlinear GCN is insufficient for capturing complex interactions. 
	
	\item Compared with NeuMF and GC-MC, SVD++ achieves higher Recall@20 and NDCG@20 on all datasets, which demonstrates that explicitly incorporating first-order neighbors into user embeddings is more effective at capturing complex interactions than nonlinear networks in NeuMF and GC-MC.
	
	\item NGCF achieves a significant improvement over GC-MC and SVD++  on the three datasets, which indicates that explicitly aggregating high-hop neighbors can effectively capture important high-order interactions and further improve model performance.
	
	\item LR-GCCF generally achieves better performance than NGCF in all cases. This might be owning to the removal of the nonlinear activation function in LR-GCCF, which can improve the learning process of GCN-based methods.
	
	\item LightGCN significantly outperforms LR-GCCF on all datasets, which illustrates that nonlinear network layers are redundant in terms of enhancing the capture of high-order collaborative signals. 
	
	\item  {SGL significantly outperforms LightGCN on the Yelp2018 and Amazon-Book datasets, which demonstrates the effectiveness of supplementing the recommendation task with self-supervised learning. Unexpectedly, SGL is slightly weaker than LightGCN on the Gowalla dataset, which may be due to the loss of important interaction information caused by SGL's utilization of node dropout and edge dropout to construct different graph views. However, the superior performance of SGL on Yelp2018 and Amazon-Book may be due to the larger number of spurious interactions in these two datasets, so the dropout and contrastive learning strategies can enhance the model robustness and thus alleviate the negative impact of spurious interactions on user preference modeling.}
	
	\item Our proposed JMP-GCF yields the best performance across all cases. Specifically, JMP-GCF outperforms LightGCN w.r.t. Recall@20 by 2.24\%, 8.17\%, and 21.41\% on the Gowalla, Yelp2018, and Amazon-Book datasets, respectively. Although JMP-GCF does not improve much on the Gowalla dataset, it substantially outperforms other methods on the Yelp2018 and Amazon-Book datasets. This is probably because multi-grained popularity features are incorporated in JMP-GCF and the users in the Yelp2018 and Amazon-Book datasets are more sensitive to popularity than those in the Gowalla dataset.  {Additionally, JMP-GCF outperforms SGL w.r.t. Recall@20 by 3.37\%, 4.00\%, and 4.39\% on the Gowalla, Yelp2018, and Amazon-Book datasets, respectively. Such results further verify the superiority of incorporating multi-grained popularity features. }
	
	\item Another observation is that the overall performance is best on the Gowalla dataset and worst on the Amazon-Book dataset. We believe that such results may be attributed to the interaction sparsity and application scenarios of the datasets. Specifically, according to Table \ref{tab:statistics}, the Amazon-book dataset has the highest sparsity, which leads to the lowest accuracy of recommendations. Although the Yelp2018 is denser than the Gowalla, the performance on the Yelp2018 is significantly weaker than that on the Gowalla. Such results may be attributed to the variability of the datasets in terms of application scenarios. Specifically, the application scenario of the Yelp2018 is restaurant and bar recommendations. Intuitively, most people tend to dine nearby, i.e., the geographic location feature in the Yelp2018 dataset is closely related to the user interaction behavior, but that feature is not taken into count in recommendation generation. In contrast, Gowalla is a check-in dataset, in which the user interactions are relatively less correlated with geographic location. We think that even though Gowalla is sparser than Yelp2018, this kind of scenario-related factors lead to better performance of Gowalla than that of Yelp2018 when only ID information is used.
\end{itemize}

To better verify the superiority of JMP-GCF over the traditional collaborative filtering baseline SVD++ and the strongest GCN-based baseline LightGCN, we investigate their performance $w.r.t$ different size $k$ of the recommendation lists. As shown in Figure \ref{fig:topk}, the curve of JMP-GCF always lies above the curves of SVD++ and LightGCN, which further indicates the effectiveness of JMP-GCF. 
In addition, we find that a phenomenon is demonstrated on all three datasets, $i.e.$, JMP-GCF achieves a greater improvement on NDCG@$k$ compared to Recall@$k$, which indicates that JMP-GCF has a stronger advantage in ranking preference modeling.

\renewcommand{\arraystretch}{2}
\renewcommand\tabcolsep{2.5 pt} 
\begin{table}
	\centering   
	\caption{Performance of Recall@20 and NDCG@20 on the Gowalla, Yelp2018, and Amazon-Book datasets w.r.t. different components in JMP-GCF.}
	\label{tab:rq2}
	\begin{tabular}{|c|c|c|c|c|c|c|}
		\hline
		\multirow{2}{*}{{{Variants}}}&
		\multicolumn{2}{c|}{{{Gowalla}}}&\multicolumn{2}{c|}{{{Yelp2018}}}&\multicolumn{2}{c|}{{{Amazon-Book}}}\cr\cline{2-7}
		&{recall}&{ndcg}&{recall}&{ndcg}&{recall}&{ndcg}\cr
		\hline
		\hline
		\makecell[c]{JMP-GCF/ \\ \{$ms$,$mp$,$se$,$ls$\}}& \makecell[c]{ 0.1800\\ (-3.9\%)}& \makecell[c]{0.1515\\ (-4.5\%)}&\makecell[c]{0.0650 \\ (-8.0\%)}&\makecell[c]{0.0540\\ (-6.9\%)}&\makecell[c]{0.0400\\ (-24.7\%)}&\makecell[c]{0.0301\\ (-28.9\%)}\cr
		\hline
		
		\makecell[c]{JMP-GCF/ \\ \{$ms$,$mp$,$se$\}}&\makecell[c]{0.1814\\ (-3.1\%)}&\makecell[c]{0.1528\\ (-3.6\%)}&\makecell[c]{0.0669\\ (-4.9\%)}&\makecell[c]{0.0553\\ (-4.3\%)}&\makecell[c]{0.0450\\ (-10.9\%)}&\makecell[c]{0.0350\\ (-10.9\%)}\cr
		\hline
		
		\makecell[c]{JMP-GCF/ \\ \{$ms$,$mp$\}}& \makecell[c]{ 0.1831\\ (-2.2\%)}& \makecell[c]{0.1545\\ (-2.5\%)}&\makecell[c]{0.0674\\ (-4.2\%)}&\makecell[c]{0.0554\\ (-4.2\%)}&\makecell[c]{0.0461\\ (-8.2\%)}&\makecell[c]{0.0359\\ (-8.1\%)}\cr
		\hline
		
		\makecell[c]{JMP-GCF/ \\ \{$ms$\}}& \makecell[c]{ 0.1842\\ (-1.6\%)}& \makecell[c]{0.1550\\ (-2.1\%)}&\makecell[c]{0.0689\\ (-1.9\%)}&\makecell[c]{0.0566\\ (-1.9\%)}&\makecell[c]{0.0489\\ (-2.0\%)}&\makecell[c]{0.0378\\ (-2.6\%)}\cr
		\hline
		{JMP-GCF}&{{0.1871}}&\makecell[c]{{0.1583} }&\makecell[c]{{ 0.0702}}&\makecell[c]{{0.0577} }&\makecell[c]{{0.0499} }&\makecell[c]{{0.0388}}\cr
		\hline
	\end{tabular}
\end{table}

\begin{figure*}
	\centering
	\subfigure[Gowalla-Recall@20.]{
		\centering
		\includegraphics[height=3.7cm,width=5.7cm]{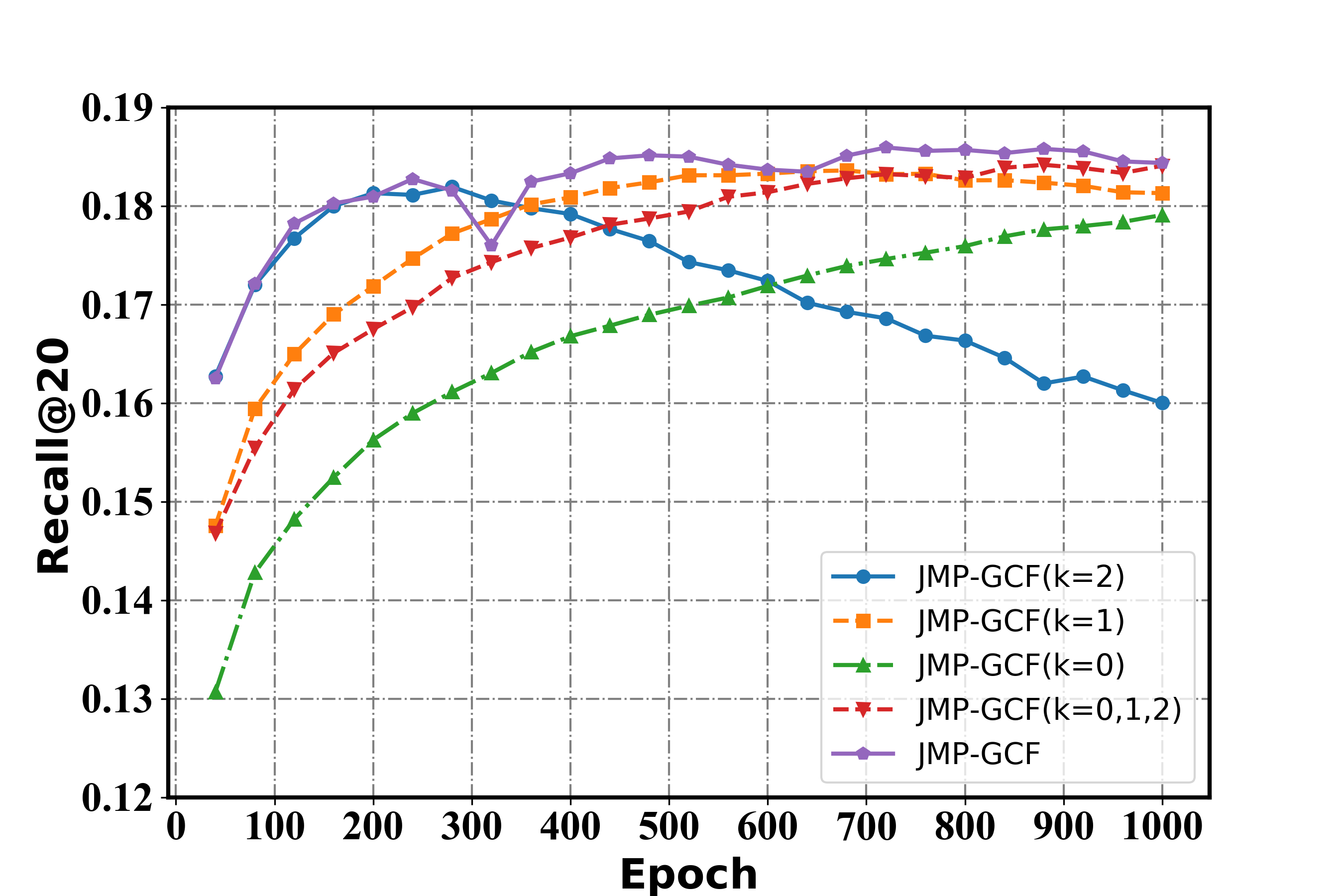}
	}%
	\subfigure[Yelp2018-Recall@20.]{
		\centering
		\includegraphics[height=3.7cm,width=5.7cm]{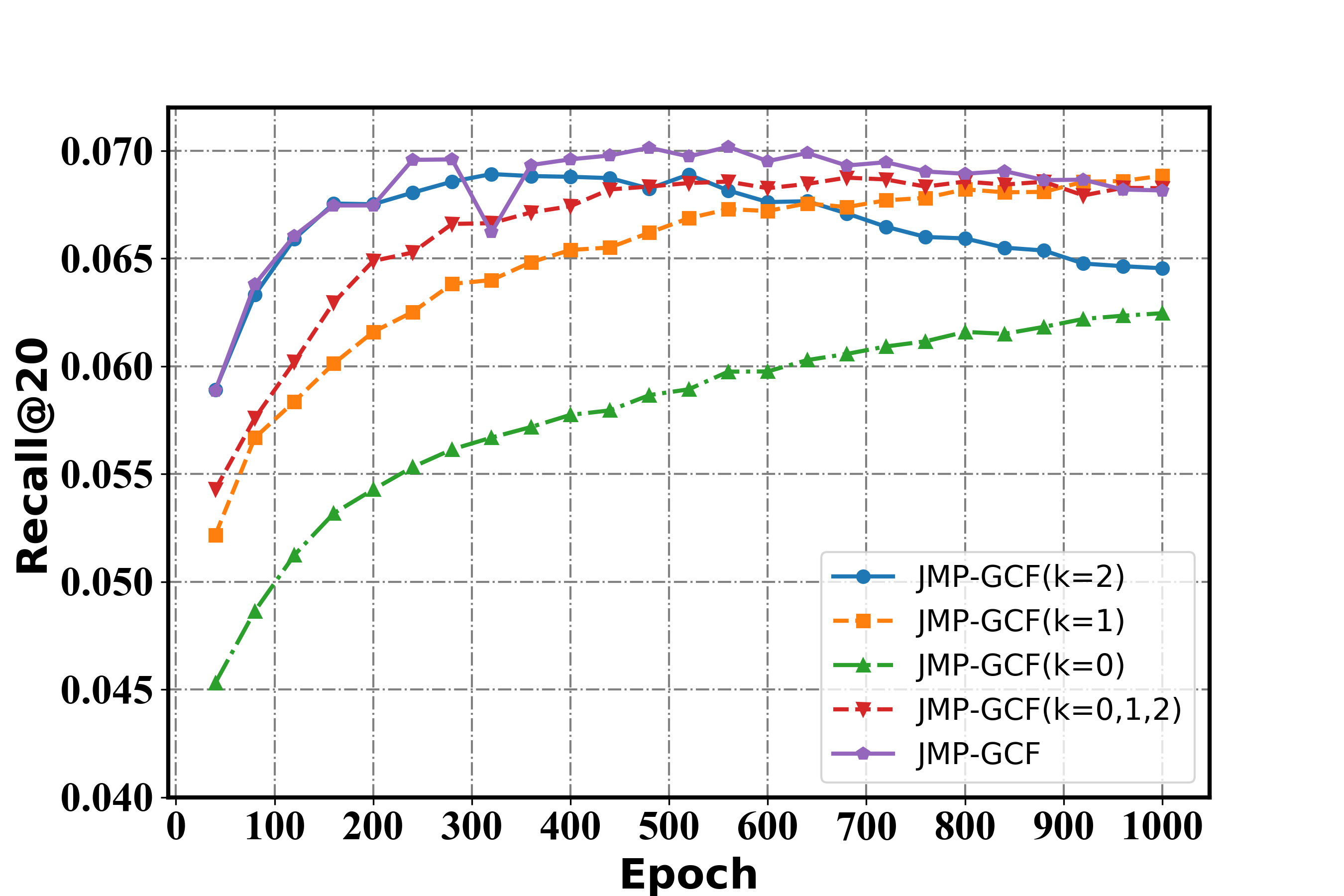}
	}%
	\subfigure[Amazon-book-Recall@20.]{
		\centering
		\includegraphics[height=3.7cm,width=5.7cm]{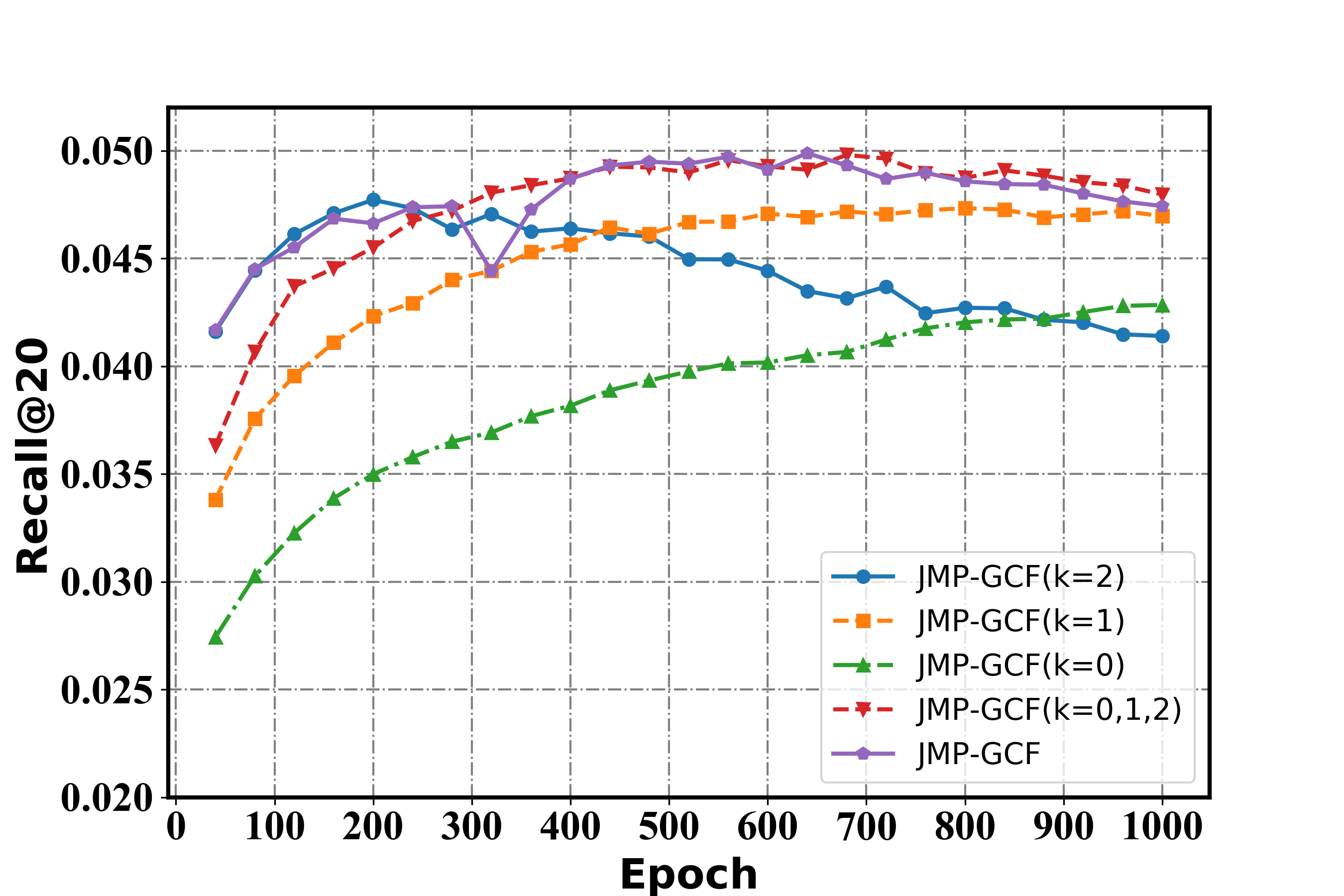}
	}%
	
	\subfigure[Gowalla-NDCG@20.]{
		\centering
		\includegraphics[height=3.7cm,width=5.7cm]{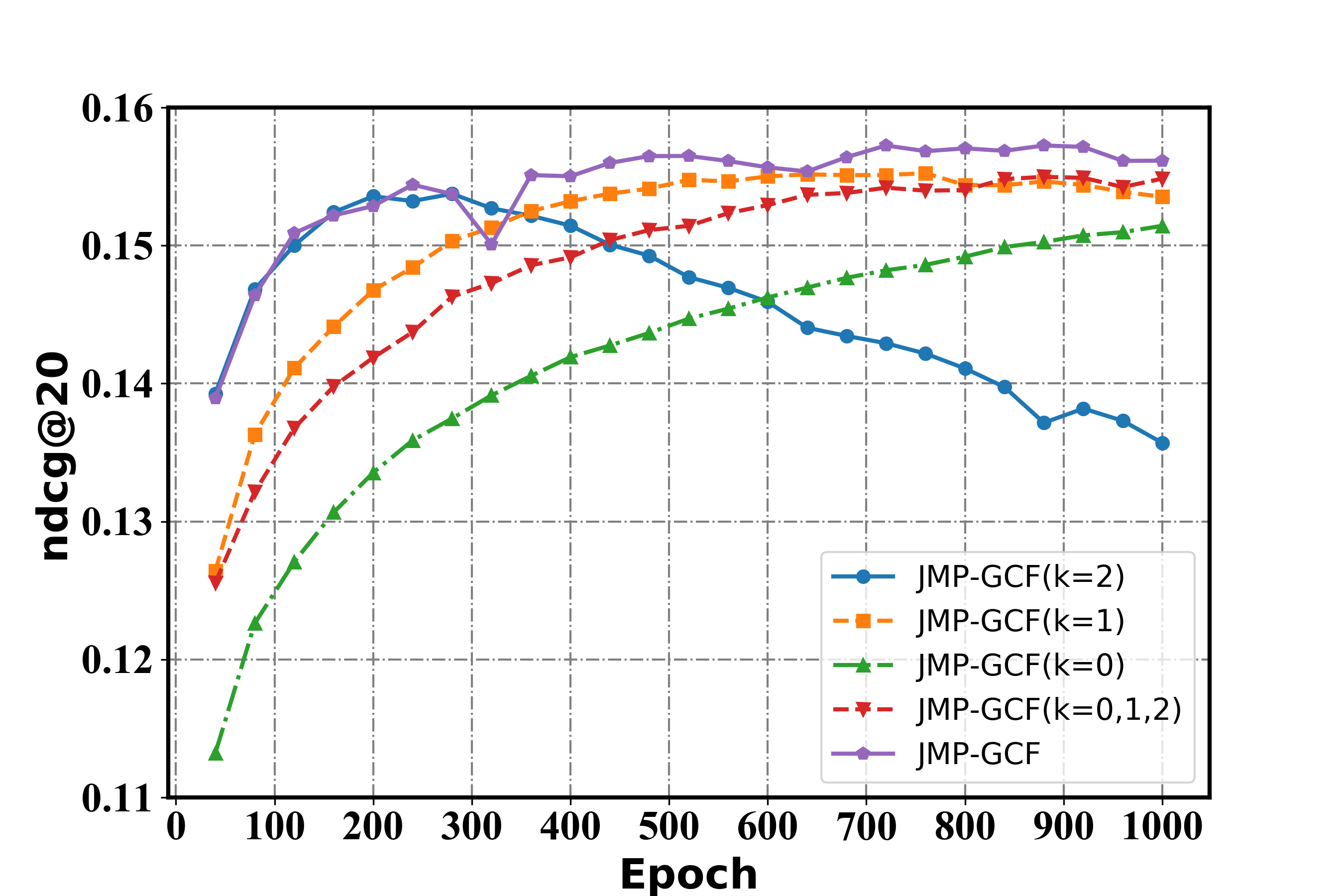}
	}%
	\subfigure[Yelp2018-NDCG@20.]{
		\centering
		\includegraphics[height=3.7cm,width=5.7cm]{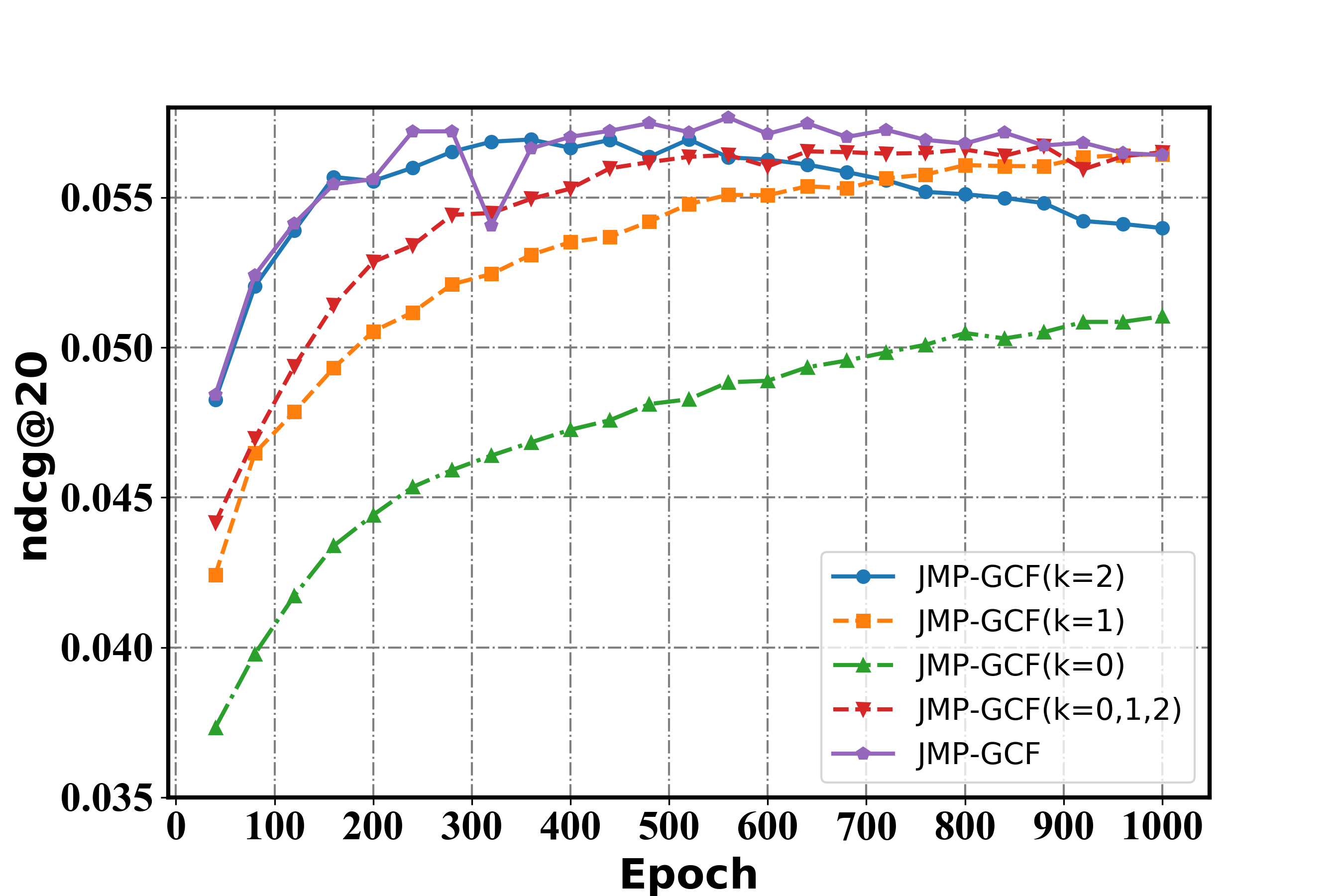}
	}%
	\subfigure[Amazon-book-NDCG@20.]{
		\centering
		\includegraphics[height=3.7cm,width=5.7cm]{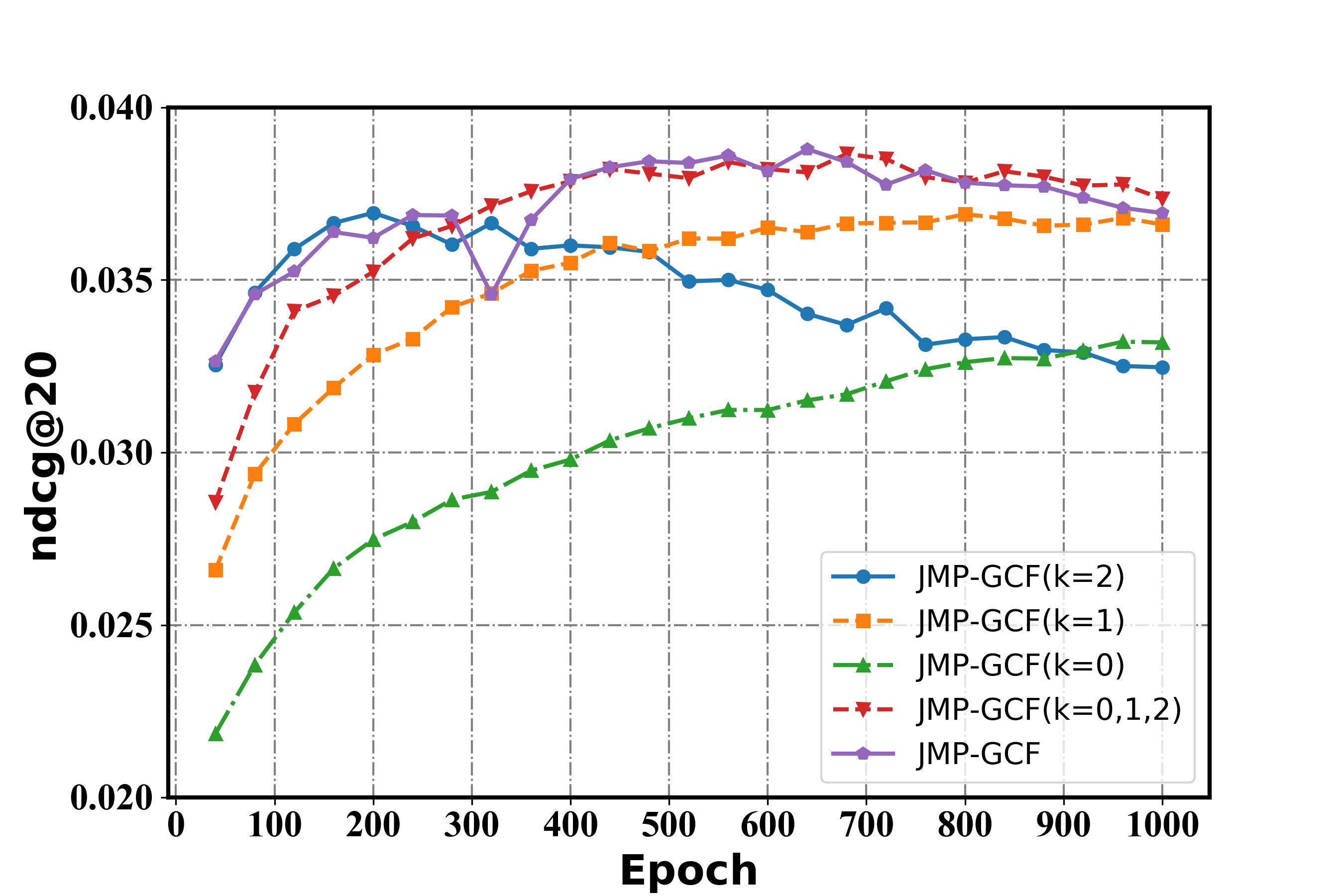}
	}%
	\caption{Performance trends of Recall@20 and NDCG@20 within 1000 training epochs w.r.t. different  granularity of popularity features on the Gowalla, Yelp2018, and Amazon-Book datasets.}
	\label{fig:popu:trend}
\end{figure*}

\subsection{Analysis of Model Components (RQ2)}
The experiments in this section are designed to validate the effectiveness of each component of JMP-GCF. 
For simplicity,  the symbol $ms$ denotes the multistage stacked training method, $mp$ denotes multi-grained popularity features, $se$ denotes separated BPR loss, and $ls$ denotes the layer selection mechanism. These symbols are used to specify the following variants of JMP-GCF.

\subsubsection{\textbf{Effect of using layer selection mechanism}} \label{subsubsec:ls}
We set up two model variants: JMP-GCF/\{$ms$,$mp$,$se$,$ls$\}, which removes $ms$, $mp$, $se$, and $ls$ from JMP-GCF; and  JMP-GCF/\{$ms$,$mp$,$se$\}, which additionally considers $ls$ on top of JMP-GCF/\{$ms$,$mp$,$se$,$ls$\}, that is, it uses the layer selection mechanism to generate the final node representations. The performance of these two variants is reported in Table \ref{tab:rq2}, which shows that the variant considering $ls$ achieves significant improvements over JMP-GCF/\{$ms$,$mp$,$se$,$ls$\} on the three datasets. This indicates that the approach of selecting optimal layers to ensure adequate capture of the layer semantics is effective.
\subsubsection{\textbf{Effect of using separated BPR loss}}\label{ex:subsubsec:loss}
We set up the model variant JMP-GCF/\{$ms$,$mp$\}, which uses $se$ instead of the traditional BPR loss used in  JMP-GCF/\{$ms$,$mp$,$se$\}, as the objective function to optimize the model parameters. The multi-grained popularity features are not considered in this variant here, $se$ only implements the joint learning of two layer semantics. The experimental results are recorded in Table \ref{tab:rq2}, which shows that JMP-GCF/\{$ms$,$mp$\} outperforms JMP-GCF/\{$ms$,$mp$, $se$\} across all cases. This indicates that the joint learning of two layer semantics is reasonable and separated BPR loss is helpful for improving model performance.
\subsubsection{\textbf{Effect of using popularity integration}}  
We set up a variant JMP-GCF/\{$ms$\} which incorporates multi-grained popularity features based on JMP-GCF/\{$ms$,$mp$\}. Specifically, in accordance with Equation \ref{eq:akeo} in Section \ref{sec:sgcb2}, we set $k$=0, $k$=1, and $k$=2 to construct the popularity features at the three granularities, and use Equation \ref{eq:pred} in Section \ref{sec:preopt} to predict the preference scores and Equation \ref{eq:loss} to jointly learn the popularity features at these three granularities. The experimental results are recorded in Table \ref{tab:rq2}, showing that JMP-GCF/\{$ms$\} achieves a significant improvement over the variant JMP-GCF/\{$ms$,$mp$\} on all three datasets. This demonstrates that popularity features play an important role in modeling user preferences and it proves that user preferences show differentiation in popularity features. In addition, we find that incorporating popularity features results in a more pronounced improvement in performance on the Yelp2018 and Amazon-Book datasets, probably because users in these two scenarios are more sensitive to popularity features.  
\subsubsection{\textbf{Effect of using multistage stacked training}} 
We compare the performance of the variant JMP-GCF/\{$ms$\} with that of JMP-GCF, which is equivalent to adding the multistage stacked training strategy to JMP-GCF/\{$ms$\}. Specifically, we use Eqs. \ref{eq:loss1} and \ref{eq:lossk} in Section \ref{subsec:mst} to implement the stacked joint learning of multi-grained popularity features. Table \ref{tab:rq2} shows that JMP-GCF outperforms JMP-GCF/\{$ms$\} in all cases, which demonstrates the effectiveness of multistage stacked training. 
It is worth mentioning that the main purpose of our design of multistage stacked training is to accelerate the model convergence, which is further verified in Section \ref{sec:rq3}.

\renewcommand{\arraystretch}{1.5}
\renewcommand\tabcolsep{1.5 pt}
\begin{table}
	\centering
	\caption{Performance of Recall@20 and NDCG@20 on the Gowalla and Yelp2018 dataset w.r.t. the selection of graph convolution layer (results on the Amazon-Book dataset have the same trend, which are removed for space).}
	\label{tab:layer}
	\begin{tabular}{|c|c|c|c|c|}
		\hline
		\multirow{2}{*}{{\makecell[c]{Layers \\ Selection}}}&
		\multicolumn{2}{c|}{{{Gowalla}}}&\multicolumn{2}{c|}{{{Yelp2018}}}\cr\cline{2-5}
		&{recall}&{ndcg}&{recall}&{ndcg}\cr
		\hline
		\hline
		{(1)}&\makecell[c]{0.1668(-8.8\%)}&\makecell[c]{0.1419(-7.7\%)}&\makecell[c]{0.0605(-10.6\%)}&\makecell[c]{0.0502(-10.2\%)}\cr
		\hline
		{(2)}&\makecell[c]{0.1776(-2.1\%)}&\makecell[c]{0.1502(-1.7\%)}&\makecell[c]{0.0641(-4.4\%)}&\makecell[c]{0.0539(-2.6\%)}\cr
		\hline
		{(3)}&\makecell[c]{0.1800(-0.8 \%)}&\makecell[c]{0.1515(-0.9\%)}&\makecell[c]{0.0664(-0.8\%)}&\makecell[c]{0.0550(-0.5\%)}\cr
		\hline
		{(4)}&\makecell[c]{0.1786(-1.6\%)}&\makecell[c]{0.1502(-1.7\%)}&\makecell[c]{0.0643(-4.0\%)}&\makecell[c]{0.0541(-2.2\%)}\cr
		\hline
		{(5)}&\makecell[c]{0.1681(-8.0\%)}&\makecell[c]{0.1418(-7.8\%)}&\makecell[c]{0.0590(-13.4\%)}&\makecell[c]{0.0492(-12.4\%)}\cr
		\hline
		{(1,2,3,4,5)}&\makecell[c]{0.1730(-4.9\%)}&\makecell[c]{0.1467(-4.2\%)}&\makecell[c]{0.0585(-14.4\%)}&\makecell[c]{0.0502(-10.2\%)}\cr
		\hline
		{(2,3)}&\makecell[c]{0.1809(-0.3\%)}&\makecell[c]{0.1524(-0.3\%)}&\makecell[c]{0.0663(-0.9\%)}&\makecell[c]{0.0548(-0.9\%)}\cr
		\hline
		{(3,4)}&\textbf{0.1814}&\textbf{0.1528}&\textbf{0.0669}&\textbf{0.0553}\cr
		\hline
		{(4,5)}&\makecell[c]{0.1798(-0.9\%)}&\makecell[c]{0.1511(-1.1\%)}&\makecell[c]{0.0637(-5.0\%)}&\makecell[c]{0.0525(-5.3\%)}\cr
		\hline
	\end{tabular}
\end{table}
\renewcommand{\arraystretch}{1.5}
\renewcommand\tabcolsep{2.5 pt}
\begin{table}
	\centering   
	\caption{Performance of Recall@20 and NDCG@20 on the Gowalla, Yelp2018, and Amazon-Book dataset w.r.t.  different  granularity of popularity features.}
	\label{tab:popu}
	\begin{tabular}{|c|c|c|c|c|c|c|}
		\hline
		\multirow{2}{*}{{{Method}}}&
		\multicolumn{2}{c|}{{{Gowalla}}}&\multicolumn{2}{c|}{{{Yelp2018}}}&\multicolumn{2}{c|}{{{Amazon-Book}}}\cr\cline{2-7}
		&{recall}&{ndcg}&{recall}&{ndcg}&{recall}&{ndcg}\cr
		\hline
		\hline	
		{JMP-GCF$_{k=0}$}& { 0.1831}& {0.1545}&{0.0674}&{0.0554}&{0.0461}&{0.0359}\cr
		\hline	
		{JMP-GCF$_{k=1}$}&{0.1836}& {0.1551}&{0.0688}&{0.0564}&{0.0473}&{0.0370}\cr
		\hline
		{JMP-GCF$_{k=2}$}&0.1820&0.1539&{0.0691}&{0.0570}&{0.0479}&{0.0372}\cr
		\hline
		{JMP-GCF$_{k=0,1,2}$}& { 0.1842}& {0.1550}&{0.0693}&{0.0571}&{0.0489}&{0.0378}\cr
		\hline
		{{JMP-GCF}}& \bf{ 0.1871}& \bf{0.1583}& \bf{0.0702}& \bf{0.0577}&{\bf{0.0499} }&{ \bf{0.0388}}\cr
		\hline
	\end{tabular}
\end{table}

\subsubsection{\textbf{Study on selection of graph convolution layers}} \label{subsubsec:layer}
Despite the effectiveness of the layer selection mechanism for improving model performance, as verified in Section \ref{subsubsec:ls}, such a simple experimental comparison cannot verify that the layer selection mechanism can output the optimal graph convolution layers. For this reason, we conduct a set of comparison experiments on the Gowalla and Yelp2018 datasets with the number of graph convolution layers as an independent variable, and using the JMP-GCF/\{$ms$,$mp$,$se$,$ls$\} as the base model. We set up the layer numbers in the range of \{1,2,3,4,5\}. Table \ref{tab:layer} records the results of performance when selecting different graph convolution layers. We have the following findings:
\begin{itemize}
	\item As the number of graph convolution layers increases, the model performance shows an increasing trend, and the model performs best in both datasets when the number of layers is 3, which illustrates the importance of capturing high-order interactions. Moreover, the model performance declines when a larger number of layers is used, which validates our point in section \ref{subsec:lsm} that too high a graph convolution layer tends to introduce noise data.
	\item  The model variant that concatenates the embeddings output from the first five graph convolution layers as the node representations performs worse than the model variant that uses the combination of the third and fourth layers This suggest that the low-layer embeddings are unnecessary for generating the node representations in a GCN because the high-layer embeddings contain complete information from all the previous layers, which is consistent with the point made in Section \ref{subsec:lsm}.
	\item The model variant that uses the combination of graph convolution layers like (3,4) outperforms the model variant using combinations (2,3) and (4,5). This indicates that the third and fourth layers are the optimal combination of graph convolution layers for the Gowalla and Yelp2018 datasets. The optimal odd and even layers output by the layer selection mechanism are also the third and fourth layers; such consistency further validates the effectiveness of the layer selection mechanism. 
\end{itemize}

\subsection{Analysis of Multi-grained Popularity Features (RQ3)}\label{sec:rq3}

The experiments in this section are designed to investigate the effect of popularity features at different granularities, as well as the method of fusing multi-grained popularity features, on the GCN model. In particular, we use Equation \ref{eq:akeo} to construct three types of popularity features with different granularities by setting $k$ to 0,1, and 2, which correspond to three model variants JMP-GCF$_{k=0}$, JMP-GCF$_{k=1}$, and JMP-GCF$_{k=2}$, respectively. When $k=2$, the embeddings contain more coarse-grained popularity features, that is, the model is more sensitive to popularity. Additionally, we create a variant JMP-GCF$_{k=0,1,2}$, which uses the Equation \ref{eq:loss} to accumulate these popularity features with different granularities to achieve the joint learning of popularity and high-order interactions. JMP-GCF is equivalent to additionally considering the multistage stacked training strategy on top of JMP-GCF$_{k=0,1,2}$, to accelerate the convergence of model training and enhance the learning process for multi-grained popularity features. Table \ref{tab:popu} records the detailed performance of these model variants on the three datasets within 2000 training epochs, and Figure \ref{fig:popu:trend} shows the trend of model performance for these variants (due to space constraints, we only show the model performance for the first 1000 training epochs in Figure \ref{fig:popu:trend}). We have the following findings:
\begin{itemize}
	\item For the scenarios that consider only a single popularity granularity (that is, $k=0$, $k=1$, or $k=2$), JMP-GCF$_{k=1}$ achieves the best performance on the Gowalla dataset, whereas JMP-GCF$_{k=2}$ performs best on the Yelp2018 and Amazon-Book datasets. This illustrates that users in different scenarios have different sensitivities to popularity features. Figure \ref{fig:popu:trend} shows that JMP-GCF$_{k=2}$ consistently converges most rapidly but then gradually decreases, whereas JMP-GCF$_{k=0}$ converges most slowly. This indicates that coarse-grained popularity features can accelerate model convergence but easily lead to overfitting.
	\item The model variant JMP-GCF$_{k=0,1,2}$, which considers popularity features with three granularities, outperforms JMP-GCF$_{k=0}$, JMP-GCF$_{k=1}$, and JMP-GCF$_{k=2}$ on the three datasets. This illustrates the importance of incorporating multi-grained popularity features for user preference modeling. Figure \ref{fig:popu:trend} shows that, in terms of convergence speed, JMP-GCF$_{k=0,1,2}$ outperforms JMP-GCF$_{k=0}$ on the Gowalla dataset and outperforms JMP-GCF$_{k=0}$ and JMP-GCF$_{k=1}$ on the Yelp2018 and Amazon-Book datasets, and there is no risk of overfitting. This indicates that jointly learning multi-grained popularity features both accelerates model convergence and avoids the problem of overfitting.
	\item JMP-GCF achieves the best performance on the three datasets, which indicates that the multistage stacked training approach can better jointly learn multi-grained popularity features.  Figure 5 shows that JMP-GCF has the fastest model convergence speed on all three datasets. It is worth mentioning that the transient drop in the trend curve of JMP-GCF at 300 epochs is caused by the stacking of untrained popularity features with a new granularity when the training phase is switched.
\end{itemize}

\subsection{Hyper-parameter Studies}
In JMP-GCF, $L_2$ regularization coefficient $\lambda$ is the most important hyper-parameter, and we study here the performance of JMP-GCF $w.r.t$ different $\lambda$ on the three datasets.

\begin{figure}
	\centering
	\includegraphics[height=4.25cm,width=8.5cm]{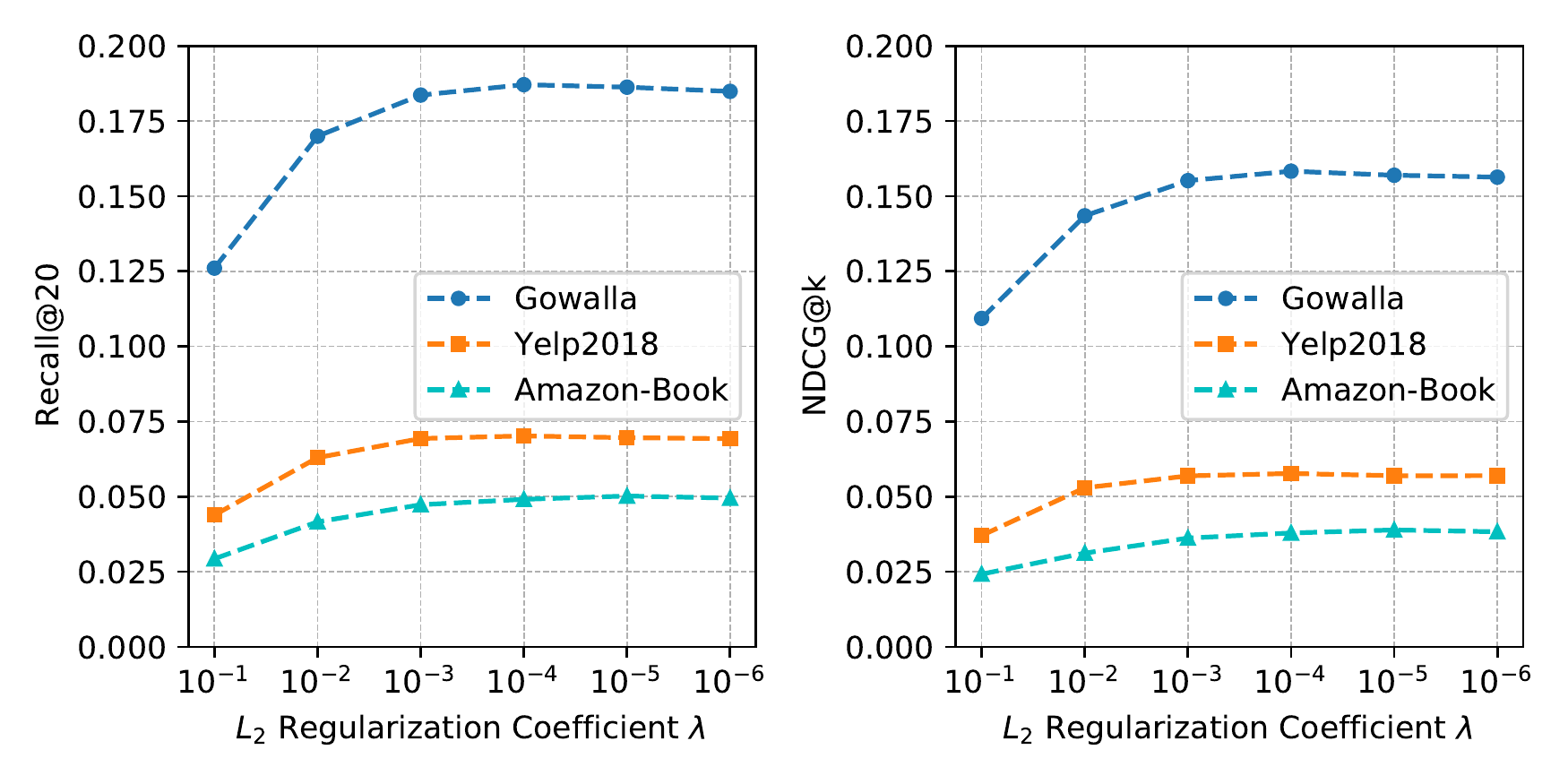}
	\caption{Performance of JMP-GCF $w.r.t.$ different $L_2$ regularization coefficient $\lambda$ on the Gowalla, Yelp2018, and Amazon-Book datasets.}
	\label{fig:l2}
\end{figure}

As shown in Figure \ref{fig:l2}, JMP-GCF achieves the best performance on the Gowalla and Yelp2018 datasets when $\lambda = 10^{-4}$, and it performs best on the Amazon-Book dataset when $\lambda = 10^{-5}$. 
More specifically, we observe that when $\lambda$ is small ($\lambda=10^{-3}$ or less), the model performance does not show significant degradation on the three datasets, which indicates that the JMP-GCF model is highly stable, $i.e.$, it does not easily fall into the issue of overfitting. 
In contrast, when $\lambda$ is set very large ($\lambda=10^{-2}$ or larger), the model consistently performs poorly, which may be due to that too strong $L_2$ regularization limits the learning ability of the model, so when applying JMP-GCF to a new dataset, we recommend not choosing too large $L_2$ regularization coefficient. 

\section{RELATED WORK}\label{sec:relatedwork}
In this section, we briefly review traditional CF methods and GCN-based methods, which are most closely related to our work.
\subsection{Traditional Collaborative Filtering Methods}
A common assumption of CF methods \cite{slim} \cite{scf} is that similar users exhibit similar preferences to items that they commonly interacted with in t8888888he past. MF \cite{mf} is the pioneering work of CF, and vectorizes users and items by mapping their ID information into embeddings, and reconstructs the interaction between users and items with the inner product of their embeddings. BiasesSVD \cite{biassvd} assumes that MF fails to accurately capture the differences in preferences among users and items, and introduces a specific bias on top of MF to compensate for embeddings' weaknesses in expressiveness.
Distinct from merely using the ID information of users and items, another type of CF methods use historical interacted items to enhance the embedding generation process. Specifically, SVD++ \cite{svd++} and FISM \cite{fism} integrate the embeddings of one-hop neighbor nodes into user representations. To explore high-order interactions and further improve model performance, HOSLIM \cite{hoslim} encodes high-hop neighbor nodes into the embeddings, but the time complexity is too high to handle the million-size datasets efficiently. 
Additionally, many researchers believe that some auxiliary properties related to users (or items), such as age, gender, occupation \cite{youtube}, multimedia features \cite{music}, and knowledge graph \cite{cke} \cite{kgnew}, are relevant to user preferences. To achieve this, SVDFeature \cite{svdfeature} effectively integrates user- and item-related attributes into the framework of MF. VBPR \cite{vbpr} and CDL \cite{cdl} use deep image feature to enhance item representations.  
In some recent efforts, deep neural networks have been applied to the CF algorithm because of heir powerful feature learning capabilities. For example, NeuMF \cite{ncf}, JNCF \cite{jncf}, and DICF \cite{dicf} use neural networks to learn nonlinear collaborative signals between users-item interactions. 

\subsection{Graph Convolution Network-Based Methods}
GCN \cite{gcn}\cite{graphsage} has received considerable attention in recent years, and is becoming increasingly important in the research field of recommendation systems. GCN-based approaches redefine the traditional rating matrix (or interaction matrix) in CF as a user-item bipartite graph, and enhance the quality of recommendations by learning local graph structural features. The traditional GCN technique consists of two main components: aggregation of neighbor nodes and feature extraction by nonlinear neural networks. To the best of our knowledge, GC-MC \cite{gcmc} was the first model to apply GCN to recommendation task. Specifically, GC-MC incorporates one-hop neighbor nodes into the embedding representations of target nodes. PinSAGE \cite{pinsage} stacks multiple graph convolution layers to capture information from high-hop neighbor nodes. NGCF \cite{ngcf} is a recommendation framework that combines GCN and MF methods, and concatenates embeddings obtained at each graph convolution layer to capture semantic information at different layers.

A recently proposed assumption about GCN is that the network layers and nonlinear activation function in GCN are redundant and easily lead to overfitting, which refers to SGCN \cite{sgc}. Inspired by SGCN, LR-GCCF \cite{lrgccf} improves upon NGCF by removing the nonlinear activation function from graph convolution layers and achieves promising improvements. Compared with LR-GCCF, LightGCN \cite{light} removes the activation function and network layers simultaneously from traditional GCN, and uses the summation of the embedding outputs from each graph convolution layer as the final representation.  {More recently, SGL \cite{sgl} incorporates a contrastive learning module into the linear GCN and jointly learns them, achieving a significant improvement over LightGCN.}

\section{Conclusion}\label{sec:conclusion}
In this work, we develop a novel recommendation model JMP-GCF, which applies the idea of joint learning to incorporate multi-grained popularity features, layer semantics, and high-order interactions into embedding generation. In addition, a multistage stacked training strategy is designed, to speed up the model convergence of JMP-GCF. We conduct extensive experiments on three public datasets and achieved the state-of-the-art performance. Further experimental analysis verified the effectiveness and rationality of each component of JMP-GCF.

In the future, we plan to further improve JMP-GCF in three main directions. First,  attention networks \cite{attention1} \cite{ acf} can be constructed to assign varying weights to different semantics and multi-grained popularity features. Second, JMP-GCF can be extended to recommendation scenarios that consider the attribute information of users and items, and these attribute features can be further divided into multiple granularities and learned jointly through the JMP-GCF framework. Finally, we intend to integrate knowledge graph \cite{cke} \cite{kgat} and causal inference \cite{ relation2} into JMP-GCF to further improve interpretability and performance.

\section*{Acknowledgment}
This work is supported by special support plan for innovation and entrepreneurship in Anhui Province.


%

%

%
%

\ifCLASSOPTIONcaptionsoff
  \newpage
\fi



%
%
%

\bibliographystyle{IEEEtran}
%

%

\begin{IEEEbiography}[{\includegraphics[width=1in,height=1.25in,clip,keepaspectratio]{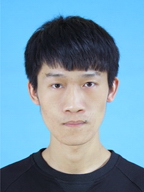}}]{Kang Liu}
	Liu Kang received his BS degree from China University of Geosciences, China, in 2017. Currently, he is working toward the MS and PhD degrees at Hefei University of Technology. His research interests include recommendation system, data mining, and multimedia analysis.
\end{IEEEbiography}

\begin{IEEEbiography}[{\includegraphics[width=1in,height=1.25in,clip,keepaspectratio]{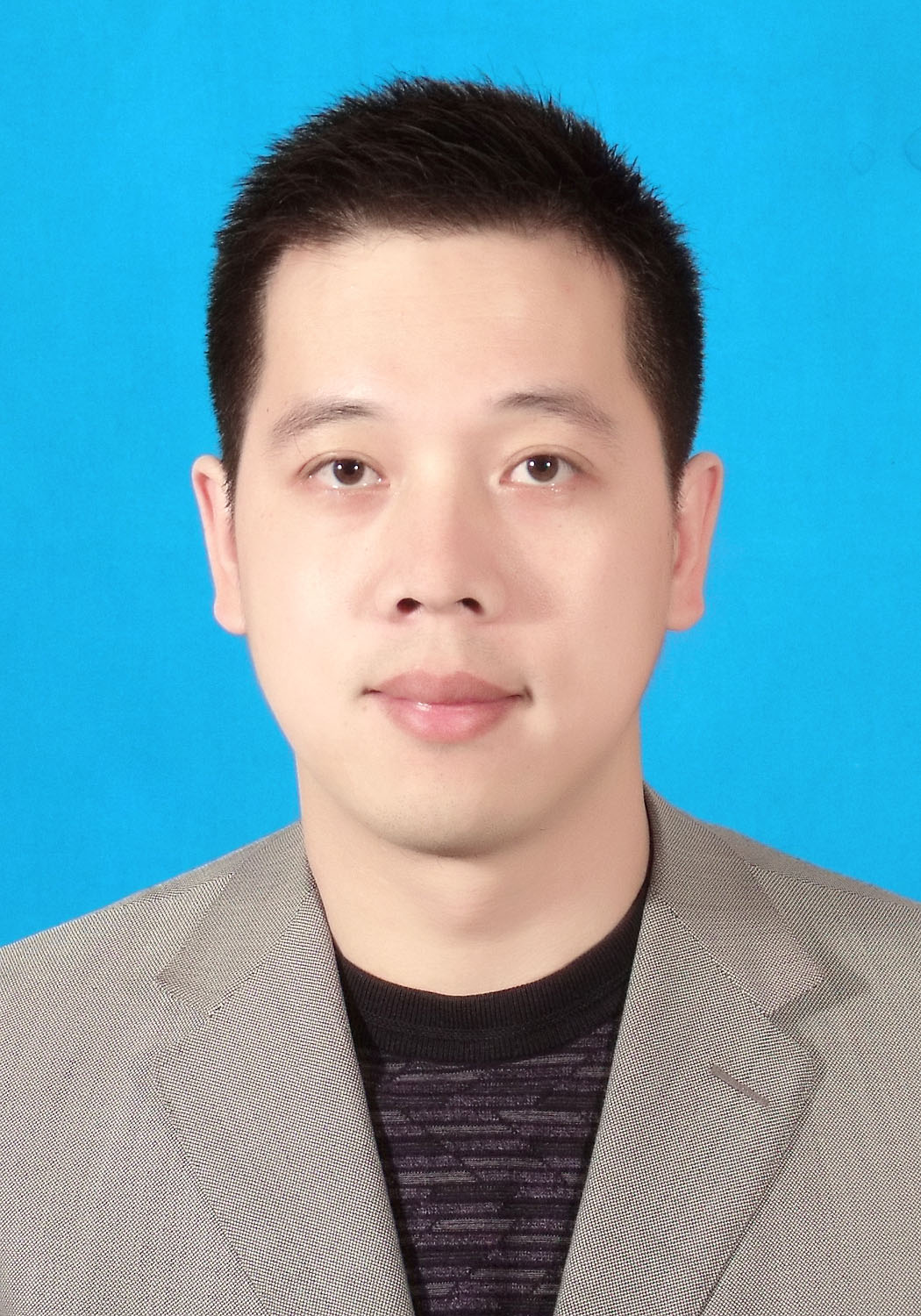}}]{Feng Xue}
	Dr. Feng Xue is a professor with the Hefei University of Technology(HFUT). He received his Ph.D. degree(June 2006) from the Dept. of Computer Science of Hefei University of Technology. His current research interests are in artificial intelligence, multimedia analysis and recommendation system.
\end{IEEEbiography}

\begin{IEEEbiography}[{\includegraphics[width=1in,height=1.25in,clip,keepaspectratio]{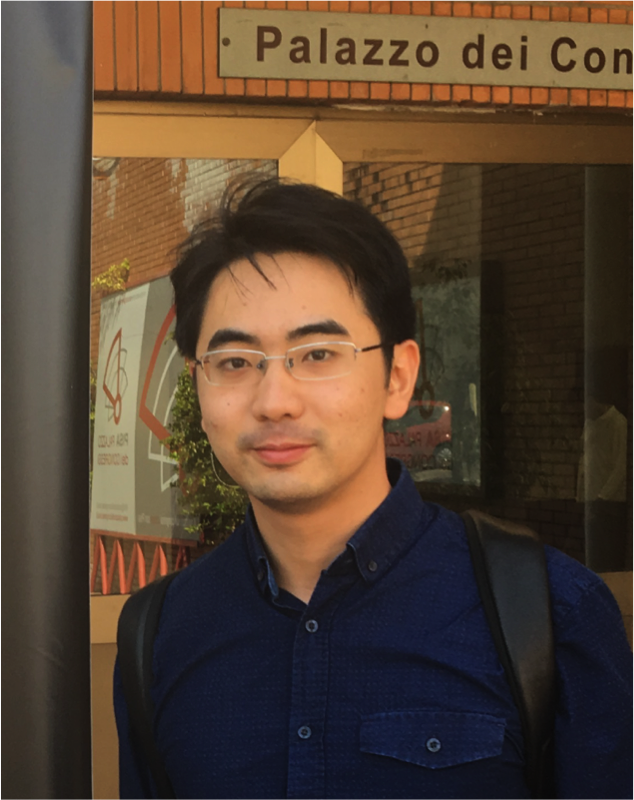}}]{Xiangnan He}
	Dr. Xiangnan He is a professor at the University of Science and Technology of China (USTC). He received his Ph.D. in Computer Science from the National University of Singapore (NUS). His research interests span information retrieval, data mining, and multi-media analytics. He has over 100 publications that appeared in top conferences such as SIGIR, WWW, and KDD, and journals including TKDE, TOIS, and TNNLS. His work has received the Best Paper Award Honorable Mention in WWW 2018 and ACM SIGIR 2016. He serves as the associate editor for ACM Transactions on Information Systems (TOIS), Frontiers in Big Data, AI Open etc. Moreover, he has served as the PC chair of CCIS 2019 and SPC/PC member for several top conferences including SIGIR, WWW, KDD, MM, WSDM, ICML etc., and the regular reviewer for journals including TKDE, TOIS, etc.
\end{IEEEbiography}

\begin{IEEEbiography}[{\includegraphics[width=1in,height=1.25in,clip,keepaspectratio]{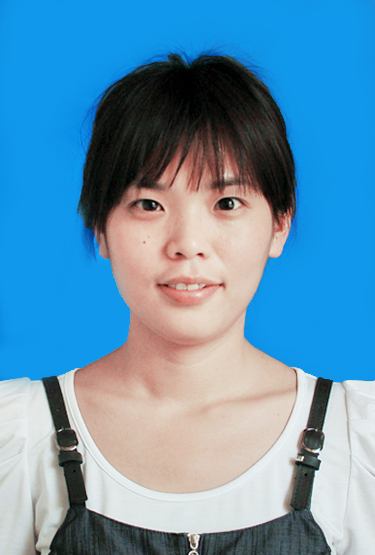}}]{Dan Guo}
	Dan Guo received the Ph.D. degree in system analysis and integration from the Huazhong University of Science and Technology, China, in 2010. She is currently an Professor at the School of Computer and Information, Hefei University of Technology. Her research interests include computer vision, machine learning, and intelligent multimedia content analysis.
\end{IEEEbiography}

\begin{IEEEbiography}[{\includegraphics[width=1in,height=1.25in,clip,keepaspectratio]{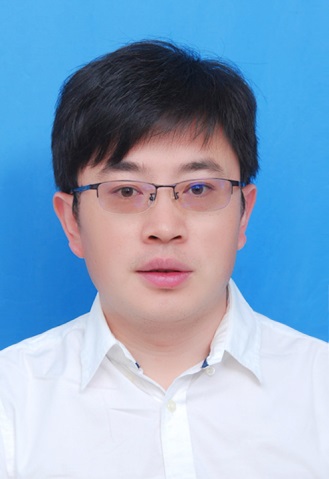}}]{Richang Hong}
	Richang Hong (M’12) received the Ph.D. degree from the University of Science and Technology of China, Hefei, China, in 2008. He is currently a Professor with the Hefei University of Technology, Hefei, China. He was a Research Fellow with the School of Computing, National University of Singapore from 2008 to 2010. He has co-authored more than 100 publications in the areas of his research interests, which include multimedia content analysis and social media. He was a recipient of the Best Paper Award in the ACM Multimedia 2010, Best Paper Award in the ACM ICMR 2015, and the Honorable Mention of the IEEE Transactions on Multimedia Best Paper Award. He served as the Associate Editor of the IEEE Multimedia Magazine, Information Sciences and Signal Processing, Elsevier and the Technical Program Chair of the MMM 2016. He is a member of ACM and the Executive Committee Member of the ACM SIGMM China Chapter.
\end{IEEEbiography}





\end{document}